\documentclass[12pt]{JHEP3}

\usepackage{epsfig}
\epsfclipon
\usepackage{multicol}
\usepackage{epsfig,bm}
\usepackage{amssymb,amsmath}
\usepackage{graphicx}
       
\newcommand{\be}{\begin{equation}}
\newcommand{\ee}{\end{equation}}
\newcommand{\bea}{\begin{eqnarray}}
\newcommand{\beas}{\begin{eqnarray*}}
\newcommand{\eea}{\end{eqnarray}}
\newcommand{\eeas}{\end{eqnarray*}}
\newcommand{\ba}{\begin{array}}
\newcommand{\ea}{\end{array}}
\def\ls{\mathrel{\lower4pt\vbox{\lineskip=0pt\baselineskip=0pt
           \hbox{$<$}\hbox{$\sim$}}}}
\def\gs{\mathrel{\lower4pt\vbox{\lineskip=0pt\baselineskip=0pt
           \hbox{$>$}\hbox{$\sim$}}}}

\def\smiley{\hbox{\large$\bigcirc$\hspace{-.80em}%
\raise.2ex\hbox{$\cdot\cdot$}\kern-.61em    
\lower.2ex\hbox{\scriptsize$\smile$}}\ }

\newcommand{\roughly}[1]{\mathrel{\raise.3ex\hbox{$#1$\kern-0.85em
\lower1ex\hbox{$\sim$}}}}

\newcommand{\lsim}{\roughly<}
\newcommand{\gsim}{\roughly>}

\def\be{\begin{equation}}
\def\beq\begin{equation}
\def\ee{\end{equation}}
\def\bea{\begin{eqnarray}}
\def\eea{\end{eqnarray}}

\def\beq{\begin{equation}}
\def\eeq{\end{equation}}
\def\beqa{\begin{eqnarray}}
\def\eeqa{\end{eqnarray}}

\newcommand{\bmat}{\left(\begin{array}}
\newcommand{\emat}{\end{array}\right)}

\title{Supermassive gravitinos, dark matter, leptogenesis and 
flat direction baryogenesis}

\author{Rouzbeh
Allahverdi$^{1}$, 
Steen Hannestad$^{2,3}$, Asko
Jokinen$^{4}$, Anupam Mazumdar$^{3,4}$, and Silvia Pascoli$^{5}$\\

${}^1$ Theory Group, TRIUMF, 4004 Wesbrook Mall, Vancouver, BC,
Canada, V6T 2A3.\\ 
$^{2}$~Department of Physics and Astronomy, University of Aarhus, 
Ny Munkegade, DK-8000 Aarhus C, Denmark.\\ 
$^{3}$~NORDITA, Blegdamsvej-17, Copenhagen, DK-2100, Denmark.\\
$^{4}$~The Niels Bohr Institute, Blegdamsvej-17, Copenhagen, DK-2100, Denmark.
\\ 
$^{5}$~Department of Physics, Theory Division, CERN, CH-1211 Geneva 23, 
Switzerland}

\abstract{In general the gravitino mass and/or the soft supersymmetry
breaking masses in the observable sector can be much larger
than the TeV scale.  Depending on the relation between the masses, new
important channels for gravitino production in the early Universe can
arise. Gravitinos with a mass above $50$ TeV decay before
big bang nucleosynthesis, which leads to relaxation of the well known 
bound on the reheating temperature
$T_{\rm R} \leq 10^{10}$~GeV. However, if the
heavy gravitinos are produced abundantly in the early Universe, their
decay can alter the abundance of the lightest supersymmetric particle. 
Moreover, they may dominate the energy density of the
Universe. Their decay will in this case increase entropy and dilute
already created baryon asymmetry and dark matter. Such considerations
put new constraints on gravitino and sfermion masses, and the reheating
temperature. In this paper we examine various cosmological
consequences of supermassive gravitinos. We discuss advnatges and
disadvantages of a large reheating temperature in connection with thermal
leptogenesis, and find that large parts of the parameter space are
opened up for the lightest right-handed (s)neutrino mass. We also
discuss the viability of Affleck-Dine baryogenesis under the constraints from 
gravitino decay, and gravitino production from the decay of Q-balls.}

\preprint{
NORDITA-2005-26.\\ TRIUMF-PP-05-04.}

\begin{document}


\section{Introduction}

Primordial inflation is the most convincing paradigm for the early
Universe~\cite{infl}. The vacuum fluctuations created during inflation
also explain the observed temperature anisotropy of the cosmic
microwave background (CMB) radiation~\cite{WMAP}. However inflation
leaves the Universe cold and void of any thermal entropy.  Entropy is
believed to be created from the decay of the coherent oscillations of
the inflaton which can happen perturbatively~\cite{Dolgov} and/or
non-perturbatively into bosons~\cite{Brandenberger,kls,chm} and
fermions~\cite{Greene}. It is necessary that the standard model (SM)
degrees of freedom are produced at this reheating stage, particularly
baryons which are required for the synthesis of light elements during
big bang nucleosynthesis (BBN) at a temperature ${\cal O}(\rm
MeV)$~\cite{bbn,subir,lowt}.

On the other hand, we do not know the full particle content of the
Universe beyond the electroweak scale, therefore we do not know what
degrees of freedom were excited right after inflation. In this respect,
supersymmetry (SUSY) acts as a building block which can explain the
hierarchy between the Planck and electroweak scales, if it is softly
broken in the observable sector at the TeV scale,
see~\cite{Nilles}. Besides its phenomenological implications, this
also has important cosmological consequences. The scalar potential of
the minimal supersymmetric standard model (MSSM) has nearly 300 flat
directions~\cite{gkm}. These flat directions can address cosmological
issues from reheating to density perturbations~\cite{Kari}. If
$R$-parity is conserved, the lightest supersymmetric particle (LSP)
will be stable and can act as a cold dark matter (CDM)
candidate. A neutralino LSP with a mass $m_{\chi} \sim 100$ GeV can
match the current observational limit when produced
thermally~\cite{Jungman}.

SUSY breaking at the TeV scale in the observable sector can be
achieved via gravity~\cite{Nilles}, gauge~\cite{gr} and
anomaly~\cite{anomaly} mediation, leading to different patterns of
supersymmetric particle masses. However, there is a priori no
fundamental reason why this scale should be favored by
nature~\footnote{String theory which is believed to be the most
fundamental theory does not provide us with a concrete answer. Rather
it provides us with a landscape with multiple vacua~\cite{Landscape},
where the SUSY breaking scale remains undetermined~\cite{Douglas},
this transcends into an uncertainty into the scale of inflation and the
required number of minimal e-foldings~\cite{Cliff0}.}.

Inspired by the string landscape~\cite{Landscape,Douglas}, there has
recently been an interesting proposal for SUSY breaking well above the
electroweak (but below the Planck) scale~\cite{split,adgr}. In this
new scheme, coined split SUSY, the masses of sfermions can be
arbitrarily larger than those of fermions.  Although such a scheme does
not attempt to address the hierarchy problem, it removes fear from
flavor changing and $CP-$violating effects induced by the light
scalars at one-loop level. Successful gauge coupling unification
requires that the gauginos be kept lighter than $100$~TeV, while
spontaneous breaking of the electroweak symmetry requires that the mass of the 
lightest Higgs be around ${\cal O}(100)$~GeV.

A priori there is no fundamental theory which fixes the scale of SUSY
breaking, but cosmological considerations can constrain it. For
example, the theory permits a very long-lived gluino. The
annihilation of gluinos alone may not efficiently reduce their
abundance below the experimental limits on the anomalous isotopes of
ordinary matter~\cite{split}. The decay of gluinos within the lifetime
of the Universe solves this potential problem, and requires
sfermions masses to be less than $10^{13}$~GeV~\cite{split,adgr}. Models
which give rise to such a {\it split} pattern of SUSY breaking masses
are typically more complicated than the conventional ones for TeV
scale SUSY breaking~\cite{split,adgr,models}. Of course, one might
also give upon supersymmetric gauge coupling unification and allow all 
supersymmetric particles to be much heavier than TeV. In this case SUSY will
be irrelevant for physics at the electroweak scale. On the other hand,
even in the context of MSSM, some of the sfermions can have a mass $\gg
1$ TeV, as happens in the so-called inverted hierarchy
models~\cite{inverted}. Therefore, under general circumstances, it is
possible that at least some of the sfermions are much heavier
than TeV.

Local SUSY naturally embeds gravity, hence supergravity, 
and implies the existence of
a new particle known as the gravitino which is the superpartner of the
graviton which is fairly long-lived. Massive gravitinos consist of
helicity $\pm 1/2$ (longitudinal) and helicity $\pm 3/2$ (transverse)
components. In the early Universe gravitinos can be produced from
thermal scatterings of gauge and gaugino quanta~\cite{thermal,gmsb},
and from the decay of sfermions~\cite{gmsb}. Gravitinos are also produced
non-thermally from the direct decay of the inflaton~\cite{nop,ajm,aem},
and from the vacuum fluctuations during the coherent oscillations of
the inflaton field~\cite{non-pert1,non-pert2,non-pert3,nps}. In the
minimal supergravity models, the gravitino mass $m_{3/2}$ is the same
as the soft breaking mass of scalars~\cite{Nilles}. However,
gravitinos can be much heavier once one goes beyond the minimal
supergravity, for example in no-scale models~\cite{no-scale}. It is
therefore possible that $m_{3/2} \gg 1$ TeV, even if SUSY is broken at
TeV scale in the observable sector.

Gravitinos with a mass below $50$~TeV decay during and/or after
BBN. Depending on the nature of the decay, there exist tight bounds
on the reheating temperature, see~\cite{cefo,kkm}. Obviously these bounds
do not apply to {\it supermassive} gravitinos, i.e. when $m_{3/2} \geq
50$~TeV. In such a case the reheating temperature could potentially be as
large as the inflaton mass, leading to  many interesting
consequences. As an example, it opens up new regions of the parameter
space for thermal leptogenesis~\cite{plumacher}, a scenario which is
sensitive to the reheating temperature as it requires the excitation of the
lightest right-handed (RH) neutrinos and their supersymmetric partners
from the thermal bath.

However other considerations can constrain the abundance of
supermassive gravitinos. Every gravitino produces one LSP upon its
decay. If the gravitino decays after the thermal freeze-out of LSPs,
then it can alter the LSP abundance. In addition, gravitinos can
dominate the energy density of the Universe if they are produced
abundantly. Entropy release from gravitino decay in this case dilutes
any generated baryon asymmetry.

However, this may turn into a virtue, for example, in the case of
Affleck--Dine (AD) baryogenesis~\cite{ad}. Depending on the parameter
space, the AD mechanism (which utilizes supersymmetric flat
directions) can generate order one baryon asymmetry. This would then
be diluted to the observed value if the gravitino decay generates
enough entropy.  Often the flat directions also fragment to form
non-topological solitons such as supersymmetric Q-balls. The Q-balls
decay slowly through their surface~\cite{cohen}, and can
themselves be a major source of late gravitino production.

In this article we consider various cosmological consequences of
models with superheavy gravitinos and/or sfermions, without delving
into model-building issues.  Such particles will be inaccessible at
future colliders, and hence cosmology will be essentially the only
window to probe and/or constrain these models.  The rest of this paper
is organized as follows. In Section II we briefly review various
sources of gravitino production and their individual contributions.
We then discuss gravitino decay and constraints from the LSP dark
matter on the abundance of supermassive gravitinos in Section III.  We
briefly review thermal leptogenesis and the effects of the gravitino
in Section IV, and identify regions of the parameter space which allow
successful leptogenesis.  Section V is devoted to supersymmetric flat
directions baryogenesis, including thermal effects and viability of
the AD mechanism in the presence of supermassive gravitinos. We also show 
that long-lived Q-balls can be a
source for copious production of gravitinos. We summarize our results
and conclude the paper in the final Section VI.

\section{Gravitino Production} 

The most important interaction terms of the gravitino field
$\psi_{\mu}$ come from its coupling to the supercurrent. In the
four-component notation, and in flat space-time, these terms are
written as (see for instance~\cite{moroi})
\beq \label{lagr}
{\cal L}_{\rm int} = -{1 \over \sqrt{2} M_{\rm P}} \partial_{\nu} X^{*} {\bar 
{{\psi}_{\mu}}} \gamma^{\nu} \gamma^{\mu} \left({1 + \gamma_5 \over 2}\right) 
\chi ~ - ~ {i \over 8 M_{\rm P}} {\bar {{\psi}_{\mu}}} 
[\gamma^{\nu},\gamma^{\rho}] \gamma^{\mu} \lambda^{(a)} F^{(a)}_{\mu \nu} 
~ + ~ {\rm h.c.}
\eeq
Here $X$ and $\chi$ denote the scalar and fermionic components of a
general chiral superfield, respectively, while $F^{(a)}_{\mu \nu}$ and
${\lambda}^{(a)}$ denote the gauge and gaugino field components of a
given vector superfield respectively.

In the limit of unbroken SUSY gravitinos are massless and the physical
degrees of freedom consist of the helicity $\pm 3/2$ (transverse)
components. After spontaneous SUSY breaking gravitino eats the
Goldstino and obtains a mass $m_{3/2}$ through the super Higgs
mechanism, and helicity $\pm 1/2$ (longitudinal) states appear as
physical degrees of freedom. When the value of $m_{3/2}$ is much smaller than 
the
momentum of the gravitino, the wave-function of the helicity $\pm 1/2$
components of the gravitino can be written as
\beq \label{goldstino}
{\psi}_{\mu} \sim i \sqrt{2 \over 3} { 1 \over m_{3/2}} \partial_{\mu} \psi,
\eeq
with $\psi$ being the Goldstino. The helicity $\pm 1/2$ states of the
gravitino will in this case essentially interact like the Goldstino
and the relevant couplings are given by an effective Lagrangian
\beq \label{efflagr}
{\cal L}_{\rm eff} = {i \over \sqrt{3} m_{3/2} M_{\rm P}} 
\left[(m^2_{X} - m^2_{\chi}) 
{X}^* {\bar \psi} \left({1 + \gamma_5 \over 2}\right) \chi ~ - ~ 
m_{\lambda} 
{\bar \psi} [{\gamma}^{\mu} , {\gamma}^{\nu}] {\lambda}^{(a)} 
F^{(a)}_{\mu \nu}\right] + {\rm h.c.}
\eeq
Here $m_{X}$ and $m_{\chi}$ denote the mass of $X$ and $\chi$ fields,
respectively, while $m_{\lambda}$ is the mass of gaugino field
$\lambda^{(a)}$. The interactions of helicity $\pm 1/2$ and helicity
$\pm 3/2$ states of the gravitino have essentially the same strength
when $\vert m_X - m_{\chi} \vert$ and $m_{\lambda}$ are smaller than
$m_{3/2}$. In the opposite limit, the rate for interactions of
helicity $\pm 1/2$ states with $X$ and $\chi$, respectively gauge and
gaugino fields, will be enhanced by a factor of $(m^2_X -
m^2_{\chi})^2/ E^2 m^2_{3/2}$ ($E$ being the typical energy involved
in the relevant process), respectively $m^2_{\lambda}/m^2_{3/2}$,
compared to that of helicity $\pm 3/2$ states.

Gravitinos are produced through various processes in the early
Universe, where the relevant couplings are given in~(\ref{lagr})
and~(\ref{efflagr}).

\begin{itemize}
\item {\it Thermal scatterings}:\\
Scatterings of gauge and gaugino quanta in the primordial thermal bath
is an important source of gravitino production, leading
to (up to logarithmic corrections)~\cite{thermal,gmsb}:
\begin{eqnarray} \label{scattering}
{\rm Helicity} \pm {1 \over 2}: \left({n_{3/2} \over s}\right)_{\rm sca} 
&\simeq & \left(1 + {M^2_{\widetilde g} \over 12 m^2_{3/2}}\right) 
\left({T_{\rm R} \over 10^{10}~{\rm GeV}}\right) \left[{228.75 \over 
g_{*}(T_{\rm R})}\right]^{3/2} 10^{-12}\,, \nonumber \\
{\rm Helicity} \pm {3 \over 2}: \left({n_{3/2} \over s}\right)_{\rm sca} 
&\simeq & \left({T_{\rm R} \over 10^{10}~{\rm GeV}}\right) \left[{228.75 \over 
g_{*}(T_{\rm R})}\right]^{3/2} 10^{-12}\,;
\end{eqnarray}
where $T_{\rm R}$ denotes the reheating temperature of the Universe,
$M_{\widetilde g}$ is the gluino mass and $g_{*}(T_{\rm R})$ is the
number of relativistic degrees of freedom in the thermal bath at
temperature $T_{\rm R}$. Note that for $M_{\widetilde g} \leq m_{3/2}$
both states have essentially the same abundance, while for
$M_{\widetilde g} \gg m_{3/2}$ production of helicity $\pm 1/2$ states
is enhanced due to their Goldstino nature.  The linear dependence of
the gravitino abundance on $T_{\rm R}$ can be understood
qualitatively. Due to the $M_{\rm P}$-suppressed couplings of the
gravitino, see~(\ref{lagr}) and~(\ref{efflagr}), the cross-section for
gravitino production is $\propto M^{-2}_{\rm P}$. The production rate
at temperature $T$ and the abundance of gravitinos produced within one
Hubble time will then be $\propto T^3$ and $\propto T$
respectively. We remind that the Hubble expansion rate at temperature
$T$ is given by $H = \sqrt{\left(g_{*} \pi^2/90 \right)} T^2/M_{\rm
P}$, where $g_*$ is the number of relativistic degrees of freedom in
the thermal bath at temperature $T$. This implies that gravitino
production from scatterings is most efficient at the highest
temperature of the radiation-dominated phase of the Universe,
i.e. when $T = T_{\rm R}$.

\item{\it Decaying sfermions}:\\ 
If $m_{3/2} < {\widetilde m}$, the decay channel $\mbox{\it sfermion} 
\rightarrow
\mbox{\it fermion} + \mbox{\it gravitino}$ is kinematically open. When 
${\widetilde m} > few
\times m_{3/2}$, the decay rate has a simple form
\begin{eqnarray} \label{decrate}
{\rm Helicity} \pm {1 \over 2}:~~~~~ \Gamma_{\mbox{\it sferm} \rightarrow 
\mbox{\it ferm} + 
\psi} &\simeq & {1 \over 48 \pi}{{\widetilde m}^5 \over m^2_{3/2}  
M^2_{\rm P}}\,, \nonumber \\
{\rm Helicity} \pm {3 \over 2}:~~~~~ \Gamma_{\mbox{\it sferm} \rightarrow 
\mbox{\it ferm} + 
\psi} &\simeq & {1 \over 48 \pi}{{\widetilde m}^3 \over M^2_{\rm P}}\,.
\end{eqnarray}
Sfermions will reach thermal equilibrium abundances, provided that
$T_{\rm R} \geq {\widetilde m}$. They promptly decay through their gauge
interactions when the temperature drops below ${\widetilde m}$.
Gravitinos are produced from sfermion decays for the whole duration
sfermions exist in the thermal bath $t \sim M_{\rm P}/{\widetilde
m}^2$. The abundance of gravitinos thus produced will then
be~\cite{gmsb}
\begin{eqnarray} \label{decay}
{\rm Helicity} \pm {1 \over 2}: \left({n_{3/2} \over s}\right)_{\rm
dec} &\simeq & \left({{\widetilde m} \over m_{3/2}}\right)^2
\left({{\widetilde m} \over 1~{\rm TeV}}\right) \left[{228.75 \over
g_{*}({\widetilde m})}\right]^{3/2} \left({N \over 46}\right) 1.2
\times 10^{-19} \,, \nonumber \\ 
{\rm Helicity} \pm {3 \over 2}: \left({n_{3/2} \over s}\right)_{\rm
dec} &\simeq & \left({{\widetilde m} \over 1~{\rm TeV}}\right)
\left[{228.75 \over g_{*}({\widetilde m})}\right]^{3/2} \left({N \over
46}\right) 1.2 \times 10^{-19}\,,
\end{eqnarray}
where $g_{*}({\widetilde m})$ is the number of relativistic degrees of
freedom at $T = {\widetilde m}$, and $N$ is the number of all sfermions
such that $m_{3/2} < {\widetilde m} < T_{\rm R}$. This result is
independent of $T_{\rm R}$, so long as $T_{\rm R} > {\widetilde m}$.  Note
that for ${\widetilde m} \gsim m_{3/2}$ gravitinos of both helicities will
be produced with approximately the same abundance.

If ${\widetilde m} \gg m_{3/2}$, helicity $\pm 1/2$ states interact with
sfermions and fermions very efficiently and can actually reach thermal
equilibrium, thus leading to $\left(n_{3/2}/s \right)_{\rm dec} \simeq
10^{-2}$. This will happen when ${\widetilde m} \geq \left(10^{4}
m^2_{3/2} M_{\rm P}\right)^{1/3}$, for example, ${\widetilde m} \geq
10^{9}$~GeV if $m_{3/2} \simeq 50$ TeV.

\item{\it Inflaton decay}:\\ 
Reheating of the Universe also leads to gravitino
production~\cite{non-pert1,non-pert2,nop,ajm} (for related studies,
see~\cite{aem,kyy,ad4}). Here we consider the case where inflaton
decays perturbatively and a radiation-dominated Universe is
established immediately after the completion of its
decay.\footnote{Full thermal equilibrium is indeed achieved very
rapidly, provided that inflaton decay products have interactions of
moderate strength. For details on thermalization,
see~\cite{thermalization}.} This in general provides a valid
description of the last stage of inflaton decay, regardless of how
fast and explosive the first stage of reheating might be due to
various non-perturbative effects~\cite{Brandenberger,kls}.

Let us denote the SUSY-conserving mass of the inflaton multiplet by
$M_{\phi}$, and the mass difference between the inflaton $\phi$ and
inflatino ${\widetilde \phi}$ by $\Delta m_{\phi}$.\footnote{The mass
difference between the inflaton and inflatino $\Delta m_{\phi}$ is in
general different from that between the standard model fermions and
sfermions ${\widetilde m}$. As an example, consider the case where the
soft breaking $({\rm mass})^2$ of both the inflaton and sfermion fields
is ${\widetilde m}^2$. We will then have $\Delta m_{\phi}
\simeq \left({\widetilde m}^{2}/2 M_{\phi} \right) \ll {\widetilde m}$ if
${\widetilde m} \ll M_{\phi}$, while $\Delta m_{\phi} \simeq {\widetilde m}$
when ${\widetilde m} \geq M_{\phi}$.} If $\Delta m_{\phi} > m_{3/2}$, the
decay $\phi \rightarrow {\widetilde \phi} + \mbox{\it gravitino}$ is 
kinematically
possible.  For $\Delta m_{\phi} > few \times m_{3/2}$, the partial
decay width will be~\cite{nop,ad4}
\begin{eqnarray} \label{phidecrate} 
{\rm Helicity} \pm {1 \over 2}: ~~~~~ \Gamma_{\phi \rightarrow 
{\widetilde \phi} + \psi} &\simeq &{1 \over 48 \pi} {(m^2_{\phi} 
- m^2_{\widetilde \phi})^4 \over
M_{\phi}^3 m^2_{3/2} M^2_{\rm P}}\, , \nonumber \\
{\rm Helicity} \pm {3 \over 2}:~~~~~ \Gamma_{\phi \rightarrow 
{\widetilde \phi} + \psi} &\simeq &{1 \over 48 \pi} {(m^2_{\phi} 
- m^2_{\widetilde \phi})^4 \over M_{\phi}^3 \Delta m^2_{\phi} M^2_{\rm P}}\, .
\end{eqnarray}
We can estimate the abundance of produced gravitinos with the help of
total inflaton decay rate $\Gamma_{\phi} = \sqrt{\left(g_{*} (T_{\rm
R}) \pi^2/90 \right)} T^2_{\rm R}/M_{\rm P}$, and the dilution factor
due to final entropy release which is given by $3 T_{\rm R}/4
M_{\phi}$.

If $\Delta m_{\phi} \ll M_{\phi}$, inflaton decay gives rise to
\begin{eqnarray} \label{phidecay} 
{\rm Helicity} \pm {1 \over 2}: \left({n_{3/2} \over s}\right)_
{\rm reh} &\simeq &\left({\Delta m_{\phi}\over m_{3/2}}\right)^2 
\left({\Delta m_{\phi}^2 
\over T_{\rm R} M_{\rm P}}\right) \left[{228.75 \over 
g_{*}(T_{\rm R})}\right]^{1/2} 1.6 \times 10^{-2}\, , \nonumber  \\
{\rm Helicity} \pm {3 \over 2}: \left({n_{3/2} \over s}\right)_
{\rm reh} &\simeq & \left({\Delta m_{\phi}^2 \over T_{\rm R} M_{\rm P}}\right) 
\left[{228.75 \over 
g_{*}(T_{\rm R})}\right]^{1/2} 1.6 \times 10^{-2}\,.
\end{eqnarray}
An interesting point is that $M_{\phi}$ drops out of the calculation,
and hence the final results in Eq.~(\ref{phidecay}) have no explicit
dependence on $M_{\phi}$. If $M_{\phi} \leq \Delta m_{\phi}$, the
gravitino abundance will be smaller than that in~(\ref{phidecay}) by a
factor of $16$. Note again that for $\Delta m_{\phi} \gsim m_{3/2}$
gravitinos of both helicities have approximately the same abundance.

Since from~(\ref{phidecay}) $n_{3/2} /s$
is inversely proportional to $T_{\rm R}$,
gravitino production from inflaton decay becomes more efficient at
lower reheating temperature. The reason is that a
smaller $T_{\rm R}$ means a smaller total decay rate $\Gamma_{\phi}$,
while the partial decay width~(\ref{phidecrate}) is independent from
$T_{\rm R}$.  Therefore decreasing $T_{\rm R}$, while suppresses the
production from thermal scatterings~(\ref{scattering}) and sfermion
decays~(\ref{decay}), can actually enhance the overall production of
gravitinos. Obviously gravitino production from inflaton decay reaches
saturation when the partial decay width equals the total decay rate
$\Gamma_{\phi}$. In this case all inflatons decay to
inflatino-gravitino pairs, and the subsequent decay of inflatinos will
reheat the Universe. In consequence, one gravitino will be produced
per inflaton quanta, resulting in a gravitino abundance
$\left(n_{3/2~}/s \right)_{\rm reh} = 3 T_{\rm R}/4 M_{\phi}$.

Gravitino production
in two-body decays of the inflaton will be kinematically
forbidden if $m_{3/2} \geq \Delta m_{\phi}$. However, the 
inflaton decay inevitably results in gravitino
production at higher orders of perturbation theory, provided that
$M_{\phi} > m_{3/2}$~\cite{ad1}. The leading order contributions come from the
diagrams describing the dominant mode of inflaton decay with gravitino
emission from the inflaton, its decay products and the decay
vertex. The partial width for inflaton decay to gravitinos is in this
case $\sim \left(M_{\phi}/M_{\rm P}\right)^2 \Gamma_{\phi}$ which,
after taking into account of the dilution factor, leads to
$\left(n_{3/2}/s \right)_{\rm reh} \sim \left(T_{\rm R}
M_{\phi}/M^2_{\rm P}\right)$. 
If we impose the bound on the inflaton mass $M_{\phi} \leq
10^{13}$~GeV from the CMB for a simple chaotic type inflation model,
and if $m_{3/2} \geq \Delta m_{\phi}$, gravitino production from
inflaton decay is subdominant compared to that of thermal
scatterings~(\ref{scattering}), and hence can be neglected.\footnote{A
similar process, non-thermal production of gravitons from inflaton
decay, can become important in models with extra
dimensions~\cite{abgp}. The large multiplicity of the Kaluza-Klein
modes of the graviton can in this case easily overcome the suppression
factor $\left(M_{\phi}/M_{\rm P}\right)^2$.}

\item{\it Non-perturbative production:}\\ 
We note that besides various perturbative production mechanisms, both of 
the helicity states can be excited non-perturbatively during the
coherent oscillations of the inflaton. This was first discussed
in~\cite{non-pert1} and then elaborated in \cite{non-pert2}. Right
after inflation the helicity $\pm 1/2$ component, i.e. the Goldstino,
is essentially the inflatino. For simple models with a single chiral 
superfield, it was
shown that this component can be produced abundantly
$\left(n_{3/2}/s \right) \leq \left(T_{\rm
R}/M_{\phi}\right)$~\cite{non-pert2}. The reason is that its
couplings, given in Eqs.~(\ref{lagr}) and~(\ref{efflagr}), are not
necessarily $M_{\rm P}$-suppressed (contrary to the helicity $\pm 3/2$
states). However, as explicitly shown in~\cite{non-pert3}, it also
decays quickly along with the inflaton through derivative interactions, and 
hence poses no danger. Realistic models include at least two chiral 
superfields such
that the inflation sector is different from the sector responsible for
SUSY breaking in the vacuum. In these models also most of
the spin-$1/2$ fermions produced during inflaton oscillations decay in
form of inflatinos, provided that the scales of inflation and
present day SUSY breaking are sufficiently separated~\cite{nps}. The
helicity $\pm 3/2$ components of the gravitino have $M_{\rm P}$ suppressed 
coupling all the time. In consequence, they are produced less abundantly 
$\left(n_{3/2}/s \right) \leq
\left(M_{\phi}/M_{\rm P}\right) \left(T_{\rm R}/M_{\rm
P}\right)$~\cite{non-pert1}, compared to the direct decay of the
inflaton, thermal scatterings and sfermion decays. We will therefore
ignore the contribution from non-perturbative production of gravitinos
in the following.

\end{itemize}

\noindent
To summarize, the total gravitino abundance is given by
\beq \label{total1}     
\left({n_{3/2} \over s}\right)= 
\left({n_{3/2} \over s}\right)_{\rm sca} + 
\left({n_{3/2} \over s}\right)_{\rm dec} + 
\left({n_{3/2} \over s}\right)_{\rm reh}.
\eeq
As long as $m_{3/2} \leq T_{\rm R}$, gravitinos are
always produced in thermal scatterings of gauge and gaugino quanta. 
In addition, sfermion and
inflaton decays also contribute to gravitino production if $m_{3/2} < 
{\widetilde m} \leq T_{\rm R}$ and $m_{3/2} < \Delta m_{\phi}$ respectively. 
Then it turns out
from~(\ref{scattering}),~(\ref{decay}) and~(\ref{phidecay}) that sfermion 
decays will
be the dominant source of gravitino production unless $T_{\rm R} > 1.2
\times \left({\widetilde m}^3/m^2_{3/2}\right)$ and/or $\Delta
m_{\phi} > 0.37 {\widetilde m}$. Hence, for $m_{3/2} < {\widetilde m}
\leq T_{\rm R}$, the most important contribution in~(\ref{total1}) in
general comes from sfermion decays (see also footnote 3 on page 6).

%
%


\section{Gravitino Decay}

\begin{itemize}
\item{\it Stable gravitino}:\\ 
First, we briefly consider the case for stable gravitinos. If
the gravitino is the LSP, and $R$-parity is conserved, it will be
absolutely stable. Its total abundance (including both helicity $\pm
1/2$ and $\pm 3/2$ states) will in this case be constrained by the
dark matter limit $\Omega_{3/2} h^2 \leq 0.129$, leading to
\begin{equation}
\left(\frac{n_{3/2}}{s}\right) \leq 
4.6 \times 10^{-10} \left({1~{\rm GeV} \over m_{\chi}}\right).
\end{equation}
This implies that the individual contributions from
Eqs.~(\ref{scattering}), ~(\ref{decay}) and~(\ref{phidecay}) should
respect this bound. As an example, consider the case with $m_{3/2} =
10$ GeV and $M_{\widetilde g} \simeq 1$ TeV. This results in the
constraints $T_{\rm R} \leq 5.5 \times 10^8$ GeV, ${\widetilde m} \leq
33$ TeV and $\Delta m_{\phi} \leq 140$ TeV.

\item{\it Unstable gravitino}:\\
An unstable gravitino decays to particle-sparticle pairs through the
couplings in~(\ref{lagr}), and the decay rate is given by~\cite{moroi}
\begin{equation}
\Gamma_{3/2} \simeq \left(N_g+\frac{N_{f}}{12}\right)
\frac{m^3_{3/2}}{32\pi M^2_{\rm P}}\,, 
\end{equation}
where $N_g$ and $N_f$ are the number of available decay channels into
gauge-gaugino and fermion-sfermion pairs respectively. The gravitino
decay is completed when $H \simeq \Gamma_{3/2}$, when the
temperature of the Universe is given by

\begin{equation} \label{dectemp}
T_{3/2} 
\simeq \left[\frac{10.75}{g_{\ast}(T_{3/2})}\right]^{1/4}
\left(\frac{m_{3/2}}{10^{5}~{\rm GeV}}\right)^{3/2}~6.8~{\rm MeV}\,.
\end{equation}
Here $g_*(T_{3/2})$ is the number of relativistic degrees of freedom
at $T_{3/2}$. If $m_{3/2} < 50$ TeV, gravitinos decay during or after
BBN~\cite{bbn} and can ruin its successful predictions for the
primordial abundance of light elements~\cite{subir}. If the gravitinos
decay radiatively, the most stringent bound $\left(n_{3/2}/s\right)
\leq 10^{-14}-10^{-12}$ arises for $m_{3/2} \simeq 100~{\rm GeV}-1$
TeV~\cite{cefo}. On the other hand, much stronger bounds are derived
if the gravitinos mainly decay through hadronic modes. In particular,
a branching ratio $\simeq 1$ requires that $\left(n_{3/2}/s\right)
\leq 10^{-16}-10^{-15}$ in the same gravitino mass range~\cite{kkm}.

To give a numerical example, consider the case when $m_{3/2} \simeq 1$
TeV. The abundance of a radiatively decaying gravitino is in this case
constrained to be $\left(n_{3/2}/s\right) \leq
10^{-12}$~\cite{cefo}. Then Eqs.~(\ref{scattering}),~(\ref{decay})
and~(\ref{phidecay}) result in the bounds $T_{\rm R} \leq 10^{10}$
GeV, ${\widetilde m} \leq 203$ TeV and $\Delta m_{\phi} < 1.1 \times
10^6$ GeV, respectively. If a TeV gravitino mainly decays into
gluon-gluino pairs, which will be the case if $m_{3/2} > M_{\widetilde
g}$, we must have $\left(n_{3/2}/s \right) \leq
10^{-16}$~\cite{kkm}. This leads to much tighter bounds $T_{\rm R}
\leq 10^6$ GeV, ${\widetilde m} \leq 9.4$ TeV and $\Delta m_{\phi}
\leq 11$ TeV.

\end{itemize}


\subsection{Decay of Supermassive Gravitinos and Dark Matter Abundance}

We now turn to {\it supermassive} gravitinos with a mass $m_{3/2} \geq
50$ TeV. If one insists on a successful supersymmetric gauge coupling
unification, the gaugino masses should be below $100$ TeV. This
implies that the gravitino will not be the LSP. The decay of
supermassive gravitinos happens sufficiently early in order not to
affect the BBN. Nevertheless, their abundance can still be constrained
due to different considerations. Gravitino decay produces one LSP per
gravitino. This non-thermal component may exceed the dark matter limit
if the decay happens below the LSP freeze-out temperature $T_{\rm
f}$. The freeze-out temperature is given by~\cite{Jungman}
\beq \label{freeze}
T_{\rm f} = {m_{\chi} \over x_{\rm f}}~,~ x_{\rm f} = 28 + {\rm ln} 
\left\{{1~{\rm TeV} \over m_{\chi}} {c \over 10^{-2}} \left[{86.25 \over 
g_*(T_{\rm f})}\right]^{1/2}\right\},
\eeq
where $m_{\chi}$ is the LSP mass and we have parameterized the
non-relativistic $\chi$ annihilation cross-section as
\beq \label{cross}
\langle \sigma_{\chi} v_{\rm rel} \rangle = {c \over m^2_{\chi}}.
\eeq
Note that neutralinos reach kinetic equilibrium with the thermal
bath, and hence become non-relativistic, very quickly at temperatures
above MeV~\cite{kamionkowski}. The exact value of $c$ depends on the
nature of $\chi$ and its interactions. For Bino-like LSP, $c$ can be
much smaller than for a Wino- or Higgsino-like one.  When sfermions
are much heavier than the neutralinos, $c = 3 \times 10^{-3}$ for a
Higgsino LSP and $c = 10^{-2}$ for a Wino LSP (including the effects
of co-annihilation)~\cite{adgr}.

Gravitino decay occurs after the LSP freeze-out if $T_{3/2} < T_{\rm
f}$, which translates into an upper bound on the gravitino mass
\beq \label{after}
m_{3/2} < \left({m_{\chi} \over 
1~{\rm TeV}}\right)^{2/3} \left[{g_{*}(T_{3/2}) \over 86.25}\right]^{1/6}~ 
4.3 \times 10^7~{\rm GeV}.
\eeq
This implies that for $m_{\chi} = 1$ TeV, gravitinos with a mass
$m_{3/2} < 4 \times 10^7$ GeV decay when thermal annihilation of LSPs
is already frozen. This decay produces one LSP per gravitino.
The dark matter limit $\Omega_{\chi} h^2 \leq 0.129$ constrains the
total LSP abundance to obey
\beq \label{dmlimit}
{n_{\chi} \over s} \leq 4.6 \times 10^{-10} \left({1~{\rm GeV} \over 
m_{\chi}}\right).
\eeq
The final abundance of LSPs produced from gravitino decay depends on
their annihilation rate. The rate of annihilation of
non-relativistic LSPs is given by
\beq \label{ann} 
\Gamma_{\chi} = \langle \sigma_{\chi} v_{\rm rel} \rangle ~ n_{\chi} = 
c {n_{\chi} \over m^2_{\chi}}. 
\eeq
If $\Gamma_{\chi} \geq \Gamma_{3/2}$, annihilation will be efficient and 
reduce the LSP abundance to 
\beq \label{lspeff}
{n_{\chi} \over s} \simeq 41.58 \left[{10^{-2} \over c}\right] 
\left[{86.25 \over g_{*}(T_{3/2})} 
\right]^{1/4} {m^2_{\chi} \over (m^3_{3/2} M_{\rm P})^{1/2}}.
\eeq
Otherwise, gravitino decay contributes an amount $n_{3/2}/s$ to the
LSP abundance. Having a large abundance of gravitinos, i.e.
$\left(n_{3/2}/s \right) > 4.6 \times 10^{-10} \left(1~{\rm
GeV}/m_{\chi}\right)$, is therefore potentially dangerous and requires
special attention.

The condition for efficient annihilation of LSPs whose abundance
$\left(n_{\chi}/s \right) \geq 4.6 \times 10^{-10} \left(1~{\rm
GeV}/m_{\chi}\right)$ at the time of gravitino decay translates into a
lower bound on the gravitino mass
\beq \label{eff}
m_{3/2} \geq \left[{10^{-2} \over c}\right]^{2/3} 
\left[86.25 \over {g_{*}(T_{3/2})}\right]^{1/6} 
\left({m_{\chi} \over 1~{\rm TeV}}\right)^{2}~ 2 \times 10^7~{\rm GeV}. 
\eeq
If $m_{3/2}$ is in the window given by~(\ref{after}) and~(\ref{eff}),
the final abundance of non-thermal LSPs will be given
by~(\ref{lspeff}). It satisfies the dark matter limit~(\ref{dmlimit}),
and can account for the CDM for those values of $m_{\chi}$ and
$m_{3/2}$ which saturate the inequality in~(\ref{eff}).

Eqs.~(\ref{after}) and~(\ref{eff}) can be simultaneously satisfied
only if
\beq \label{effcond}
m_{\chi} \leq \left[{c \over 10^{-2}}\right]^{1/2} 
\left[{g_{*}(T_{3/2}) \over 86.25}\right]^{1/4} ~ 1.8~{\rm TeV}.
\eeq
As a matter of fact, this is also the condition such that thermal
abundance of LSPs at freeze-out respects the dark matter limit. It is
not surprising as $T_{3/2} = T_{\rm f}$ when the inequality
in~(\ref{effcond}) is saturated. For the saturation value of
$m_{\chi}$ thermal LSP abundance gives rise to $\Omega_{\chi} h^2 =
0.129$. For smaller $m_{\chi}$ the thermal component is not
sufficient, while for larger $m_{\chi}$ thermal LSPs overclose the
Universe.

For values of $m_{\chi}$ respecting the bound in~(\ref{effcond}),
there always exists a window for $m_{3/2}$ such that gravitinos decay
after the freeze-out while at the same time non-thermal LSPs
efficiently annihilate and their abundance respects the dark matter
limit~(\ref{dmlimit}). In this mass window the abundance of thermal
LSPs is too low to account for dark matter. On the other hand,
non-thermal dark matter will be a viable scenario when the inequality
in Eq.~(\ref{eff}) is saturated (for non-thermal production of LSP dark
matter from gravitino decay, see also~\cite{adgr,ggw}). The gravitino mass
window becomes narrower as $m_{\chi}$ increases. It will eventually
disappear when $m_{\chi}$ reaches the upper bound
in~(\ref{effcond}). 
For the canonical choice of $c = 10^{-2}$, the gravitino mass window
shrinks to a single point $m_{3/2} = 6.3 \times 10^7$ GeV at the
saturation value $m_{\chi} = 1.8$ TeV.  For larger LSP masses it is
necessary that gravitinos which decay after the freeze-out are not
overproduced, i.e. that $\left(n_{3/2}/s \right) \leq 4.6 \times
10^{-10} \left(1~{\rm GeV}/m_{\chi}\right)$. Otherwise, gravitinos
must decay above the freeze-out temperature, i.e. the opposite
inequality as in~(\ref{after}) must be satisfied. Then gravitino decay
does not affect the final LSP abundance as $T_{3/2} > T_{\rm
f}$. However, for masses violating the bound in~(\ref{effcond})
thermal LSPs overclose the Universe. Therefore a viable scenario of
LSP dark matter in this case requires late entropy generation.

\subsection{Gravitino Non-domination}

We now consider the constraints from dark matter abundance on
gravitino decay in more detail. Assuming that there is no other stage
of entropy generation, the Universe will remain in the
radiation-dominated phase after reheating. During this period the
scale factor the Universe increases as $a \propto
H^{-1/2}$. Gravitinos become non-relativistic at
\beq \label{nonrel}
H_{\rm non} \simeq \left({m_{3/2} \over E_{\rm p}}\right)^2 H_{\rm p},
\eeq
where $H_{\rm p}$ denotes the expansion rate when (most of the)
gravitinos are produced and $E_{\rm p}$ is the energy of gravitinos
upon their production. If thermal scatterings are the main source of
gravitino production, $H_{\rm p} \sim T^2_{\rm R}/M_{\rm P}$ and
$E_{\rm p} \sim T_{\rm R}$. On the other hand, if sfermion decays
dominate gravitino production, $H_{\rm p} \sim {\widetilde m}^2/M_{\rm
P}$ and $E_{\rm p} = {\widetilde m}/2$. Finally, if most of the
gravitinos are produced in inflaton decay, $H_{\rm p} \sim T^2_{\rm
R}/M_{\rm P}$ and $E_{\rm p} \simeq \Delta m_{\phi}$.

For $H < H_{\rm p}$ the energy density of the gravitinos is redshifted
$\propto a^{-3}$, compared to $\propto a^{-4}$ for
radiation. Initially the gravitino energy density is $\rho_{3/2} = n_{3/2}
E_{\rm P}$, while the energy density in radiation is $\rho_{\rm rad} =
\left(\pi^2/30 \right) g_{*}(T_{\rm p}) T^4_{\rm p}$. Gravitinos will
dominate when $\rho_{3/2}/\rho_{\rm rad}$ is compensated by the slower
redshift of $\rho_{3/2}$. This happens at
\beq \label{domhubble}
H_{\rm dom} \simeq {16 \over 9} \left({n_{3/2} \over s}\right)^2 
\left({E_{\rm p} \over 
T_{\rm p}}\right)^2 H_{\rm non} = 8.9 
\left[{g_{*}(T_{\rm p}) \over 228.75}\right]^{1/2} \left({n_{3/2} 
\over s}\right)^2 {m^2_{3/2} \over M_{\rm P}},
\eeq
where we have used $s = \left(2 \pi^2 /45 \right) g_{*}(T_{\rm p})
T^3_{\rm p}$. Here $g_{*}(T_{\rm p})$ is the number of relativistic
degrees of freedom in the thermal bath at the temperature $T_{\rm p}$
when gravitinos are produced. Gravitino non-domination therefore
requires that $H_{\rm dom} < \Gamma_{3/2}$, i.e. that gravitinos
decay while their energy density is subdominant. This translates into
the bound
\beq \label{nodom}
{n_{3/2} \over s} < \left[{228.75 \over g_{*}(T_{\rm p})}\right]^{1/4} 
\left({m_{3/2} 
\over 10^5~{\rm GeV}}\right)^{1/2} ~ 2.4 \times 10^{-8}\,,
\eeq
on the gravitino abundance.

It is seen that for $m_{3/2} \geq 50$ TeV, gravitinos will dominate if
$\left(n_{3/2}/s \right) > 1.7 \times 10^{-8}$. Thermal scatterings
alone, see (\ref{scattering}), can yield such large abundances for
extremely large reheating temperatures $T_{\rm R} > 10^{14}$ GeV (we
consider $M_{\widetilde g} \leq 100$ TeV here). Therefore they do not
lead to gravitino domination in general. The sfermion and inflaton
decays, however, can produce a sufficiently large number of gravitinos
for much lower $T_{\rm R}$. As mentioned earlier, see the discussion
after Eq.~(\ref{total1}), sfermion decays are usually the dominant
source of gravitino production when $T_{\rm R} \geq {\widetilde
m}$. We therefore concentrate on the sfermions here.

Sfermion decays, see~(\ref{decay}), will not lead to gravitino
domination if
\beq \label{nosfermdom} 
{\widetilde m} < \left({m_{3/2} \over 10^5~{\rm GeV}}\right)^{5/6} 
1.3 \times 10^8~{\rm GeV}\,.
\eeq
Here we have taken $g_{*}({\widetilde m}) = 228.75$ and $N = 46$
in~(\ref{decay}).

A successful scenario with gravitino non-domination should take the
constraints from LSP production into account. Fig. (1) depicts
different regions in the ${\widetilde m}-m_{3/2}$ plane for the choice
$c = 10^{-2}$ and $m_{\chi} = 100$ GeV. Above the solid line
gravitinos dominate, i.e. the opposite inequality as
in~(\ref{nosfermdom}) is satisfied, and hence excluded. The region
between the solid and dashed lines is defined by
\beq \label{nooversferm}
\left({1~{\rm GeV} \over m_{\chi}}\right)^{1/3} \left({m_{3/2} \over 
10^5~{\rm GeV}}\right)^{2/3} 3.4 \times 10^7~{\rm GeV} < {\widetilde m} 
< \left({m_{3/2} \over 10^5~{\rm GeV}}\right)^{5/6} 
1.3 \times 10^8~{\rm GeV}\,.
\eeq
In this region $\left(n_{3/2}/s \right) > 4.6 \times 10^{-10}
\left(1~{\rm GeV}/m_{\chi}\right)$, thus the density of LSPs produced
in gravitino decay exceeds the dark matter limit. The dotted and
dot-dashed vertical lines correspond to Eqs.~(\ref{after})
and~(\ref{eff}) respectively. The regions in black color are excluded since 
either gravitinos dominate or gravitino decay produces
too many LSPs which do not sufficiently annihilate. In region $1$ gravitinos 
decay after the
freeze-out but efficient annihilation reduces the abundance of
produced LSPs below the dark matter limit. Gravitino decay occurs
before the freeze-out in region 2 and does not affect the final LSP
abundance. Below the dashed
line sfermion decays do not overproduce gravitinos. In fact, below the
${\widetilde m} = m_{3/2}$ line such decays are kinematically
forbidden altogether. In this part of the ${\widetilde m}-m_{3/2}$
plane one has to worry about thermal scatterings though. If $T_{\rm R}
\geq 4.6 \times 10^{10} \left(1~{\rm GeV}/m_{\chi}\right)$ GeV,
scatterings will overproduce gravitinos. Therefore regions 3, 4 and 5 will not
be acceptable in this case, due to inefficient LSP annihilation.

For the values of $c$ and $m_{\chi}$ chosen in this plot, thermal
abundance of LSPs at freeze-out is too small to account for dark
matter. Non-thermal LSP dark matter from gravitino decay will in this
case be a viable scenario along the dot-dashed line.

\begin{figure}[htb]
\vspace*{-0.0cm}
\begin{center}
\epsfysize=9truecm\epsfbox{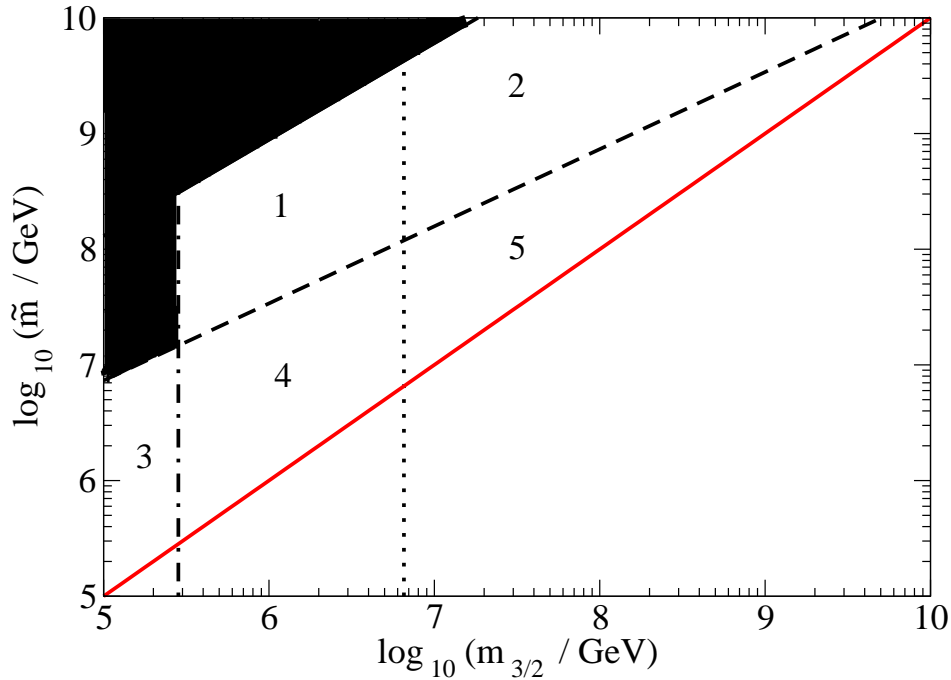}
\vspace{0.0truecm}
\end{center}
\caption{Constraints from LSP production in gravitino non-domination
case for $c = 10^{-2}$ and $m_{\chi} = 100$ GeV. The solid and dashed
lines represent the upper and lower limits in Eq.~(\protect\ref{nooversferm})
respectively. The dotted and dot-dashed lines correspond to
Eqs.~(\protect\ref{after}) and~(\protect\ref{eff}) respectively. The regions 
in black color are excluded. The solid red line is given by
$\widetilde m =m_{3/2}$.}
\end{figure}

\begin{figure}[htb]
\vspace*{-0.0cm}
\begin{center}
\epsfysize=9truecm\epsfbox{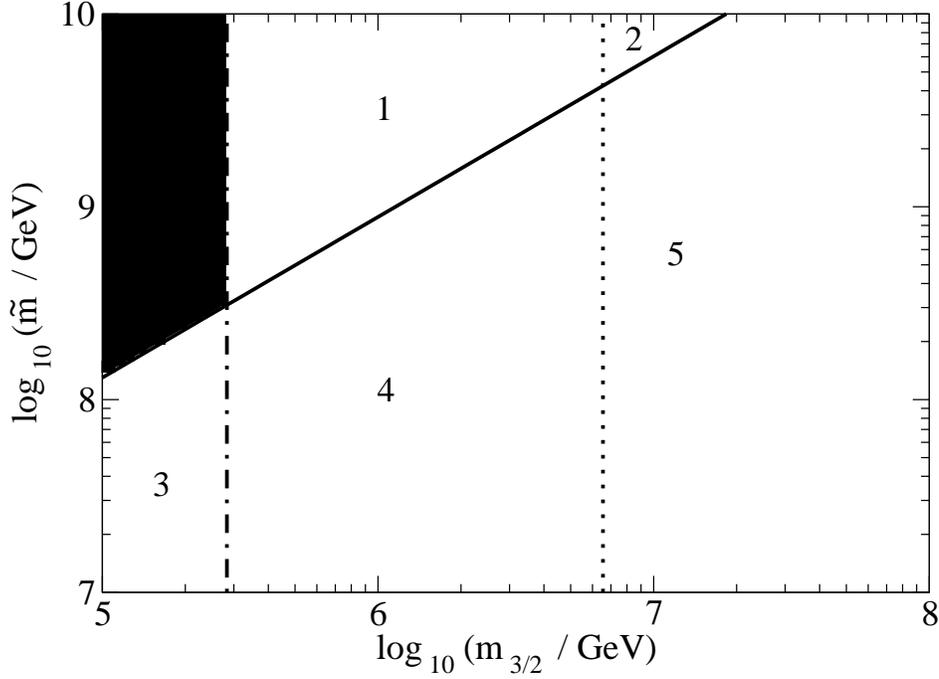}
\vspace{0.0truecm}
\end{center}
\caption{Parameter constraints from LSP production in
gravitino-dominated case for $c = 10^{-2}$ and $m_{\chi} = 100$
GeV. The dotted and dot-dashed lines correspond to Eqs.~(\protect\ref{after})
and~(\protect\ref{eff}) respectively. The region in black color
is excluded.}
\end{figure}

\subsection{Gravitino-dominated Universe}

Gravitinos eventually dominate the energy density of the Universe if
$\Gamma_{3/2} \leq H_{\rm dom}$, which happens for
\beq \label{dom}
{n_{3/2} \over s} \geq \left[{228.75 \over g_{*}(T_{\rm p})}\right]^{1/4} 
\left({m_{3/2} 
\over 10^5~{\rm GeV}}\right)^{1/2} ~ 2.4 \times 10^{-8}.
\eeq
The scale factor of the Universe $a \propto H^{-2/3}$ in the interval
$\Gamma_{3/2} \leq H < H_{\rm dom}$. Gravitino decay will then
increase the entropy density by the factor $d = \left(g_*(T_{\rm
a})T^3_{\rm a}/g_*(T_{\rm b})T^3_{\rm b}\right)$.\footnote{To be more 
precise, the dilution factor is given by $1+d$. The two definitions are 
essentially equivalent when $d \ll 1$. Obviously there is no 
gravitino dominaiton, and hence no dilution, when $d < 1$.} Here $T_{\rm
a},~T_{\rm b}$ denote the temperature of the thermal bath before and
after gravitino decay, respectively, while $g_*(T_{\rm a}),~g_*(T_{\rm
b})$ are the number of relativistic degrees of freedom at $T_{\rm a}$
and $T_{\rm b}$, respectively. Note that $d = \left(g_*(T_{\rm a})
\rho^3_{3/2}/g_*(T_{\rm b}) \rho^3_{\rm R}\right)^{1/4}$, with
$\rho_{3/2}$ and $\rho_{\rm R}$ being the energy density in the
gravitinos and radiation, respectively, at the time of gravitino
decay. The dilution factor $d$ is therefore given by
\beq \label{dilution}
d = \left({H_{\rm dom} \over \Gamma_{3/2}}\right)^{1/2} 
\left({g_*(T_{\rm a}) \over g_*(T_{\rm b})}\right)^{1/4} 
\simeq \left[{g_{*}(T_{\rm P}) \over 228.75}\right]^{1/4} 
\left({10^5~{\rm GeV} \over m_{3/2}}\right)^{1/2} 
\left[{(n_{3/2}/s) \over 2.4 \times 10^{-8}}\right]\,.
\eeq
Here we have taken $\left(g_*(T_{\rm a})/g_*(T_{\rm b}\right)^{1/4}
\simeq 1$, which is a good approximation since in a wide range $1~{\rm
Mev} \leq T \leq {\widetilde m}$ the number of relativistic degrees of
freedom $g_*$ changes between $10.75$ and $228.75$. To be more precise, the 
dilution factor is given by $1+d$. Obviously for $d < 1$ there is no 
gravitino domination, and hence no dilution. As mentioned
before, sfermions are usually the main source of gravitino production
for $T_{\rm R} \geq {\widetilde m}$, and hence we concentrate on them
here.

Gravitinos produced in sfermion decays dominate the Universe if
\beq \label{sfermdom} 
\left({m_{3/2} \over 10^5~{\rm GeV}}\right)^{5/6} 
1.3 \times 10^8~{\rm GeV} \leq {\widetilde m} \leq T_{\rm R}\,,
\eeq
in which case the dilution factor is given by
\beq \label{sfermdilution} 
d \simeq \left({10^5~{\rm GeV} \over
m_{3/2}}\right)^{5/2} \left({{\widetilde m} \over {1.3 \times
10^8~{\rm GeV}}}\right)^3\,.  
\eeq
Gravitino decay dilutes the existing LSP abundance by a factor of $d$,
while at the same time producing one LSP per gravitino. Note that
$\left(n_{3/2}/s \right) \leq 10^{-2}$ as gravitinos at most reach
thermal equilibrium when ${\widetilde m} \gg
m_{3/2}$. Eq.~(\ref{dilution}) then implies that $d \leq 5.9 \times
10^5$.
It is evident from~(\ref{dom}) that the abundance of non-thermal LSPs
upon their production from the decay of supermassive gravitinos exceeds
the dark matter limit by several orders of magnitude. Therefore any
acceptable scenario with gravitino domination requires that the LSP
annihilation be efficient at the temperature $T_{3/2}$, i.e.
that~(\ref{eff}) be satisfied.

Fig.~(2) depicts different regions in the ${\widetilde m}-m_{3/2}$
plane for the choice $c = 10^{-2}$ and $m_{\chi} = 100$
GeV. Gravitinos dominate in the region above the solid line
corresponding to Eq.~(\ref{sfermdom}). Therefore regions below this line are 
irrelevant. To the left of the dotted line,
which represents Eq.~(\ref{after}), gravitino decay occurs after the
freeze-out. To the right of the dot-dashed line, corresponding to
Eq.~(\ref{eff}), LSPs produced from gravitino decay efficiently
annihilate. The region in black color is excluded due to inefficient LSP 
annihilation. Gravitinos decay after the freeze-out in region 1, but the final
abundance of LSPs respects the dark matter limit. On the other hand,
$T_{3/2} \geq T_{\rm f}$ in region 2, and hence gravitino decay does
not affect the LSP abundance. For the values of $c$ and $m_{\chi}$
chosen here, thermal abundance of LSPs at freeze-out is too low to
account for dark matter. However, non-thermal LSP dark matter from
gravitino decay is successful along the dot-dashed line.

In passing we note that the same discussions also apply when $T_{\rm
R} < {\widetilde m}$. In this case, however, the inflaton decay will
be the main source of gravitino production as sfermions are not
excited by the thermal bath. Constraints from efficient LSP annihilation
then lead to plots similar to those in Figs.~(1),~(2), with the
${\widetilde m} - m_{3/2}$ plane replaced by the $\Delta m_{\phi} -
m_{3/2}$ plane.

Entropy release from gravitino decay also dilutes any previously
generated baryon asymmetry. This implies that baryogenesis should
either take place after gravitino decay, or generates an asymmetry in
excess of the observed value by the dilution factor given in
(\ref{dilution}). It is seen from~(\ref{dectemp}) that $T_{3/2} \leq
100$ GeV unless $m_{3/2} > 10^8$ GeV. This implies that successful
baryogenesis after gravitino decay will be possible only if gravitinos
are extremely heavy. Therefore the more likely scenario in a
gravitino-dominated Universe is generating a sufficiently large baryon
asymmetry at early times which will be subsequently diluted by
gravitino decay. We will discuss leptogenesis and baryogenesis, and the 
effect of gravitinos in detail in the next Sections.

One might invoke an intermediate stage of entropy release by the late
decay of some scalar condensate (beside inflaton) to prevent gravitino
domination. We shall notice, however, that any such decay will itself
produce gravitinos with an abundance which is inversely proportional
to the new (and lower) reheating temperature, see~(\ref{phidecay}).
\footnote{If the scalar field does not dominate the Universe, the
expression in~(\ref{phidecay}) should be multiplied by the fraction
$r$ of the total energy density $r$ which it carries.} This implies
that any stage of reheating, while diluting gravitinos which are
produced during the previous stage(s), can indeed produce more
gravitinos. Therefore entropy generation via scalar field decay is in general 
not a helpful way to avoid a gravitino-dominated Universe.

One comment is in order before closing this subsection. In both of the 
gravitino non-domination and domination
scenarios, having an LSP abundance in agreement with the dark matter
limit constrains its mass through Eq.~(\ref{effcond}). Heavier LSPs
overclose the Universe in one way or another. If $T_{3/2} \geq T_{\rm
f}$, gravitino decay will be irrelevant but thermal abundance of LSPs
will be too high. If $T_{3/2} < T_{\rm f}$, gravitino decay can in addition 
make an unacceptably large non-thermal contribution. In case of gravitino 
domination gravitino decay dilutes thermal LSPs. However, according 
to~(\ref{dom}), the decay itself overproduces non-thermal LSPs
which will not sufficiently annihilate. Therefore gravitino
domination cannot rescue a scenario with thermally overproduced
LSPs. Indeed, for $m_{\chi} \gg 1$ gravitinos should never dominate the
Universe. The problem can be solved if $T_{\rm R} < T_{\rm f}$, or if another 
stage of entropy release below $T_{\rm f}$ dilutes thermal LSPs. In both 
case, however, reheating can overproduce gravitinos and the subsequent 
gravitino decay may lead to non-thermal overproduction of LSPs. If $R$-parity 
is broken, the LSP will be unstable and its abundance 
will not be subject to the dark matter bound. Obviously its thermal and/or 
non-thermal overproduction will not pose a danger in this case.

\subsection{Solving the Boltzmann Equation}

\begin{figure}[htb]
\vspace*{-0.0cm}
\begin{center}
\epsfysize=14truecm\epsfbox{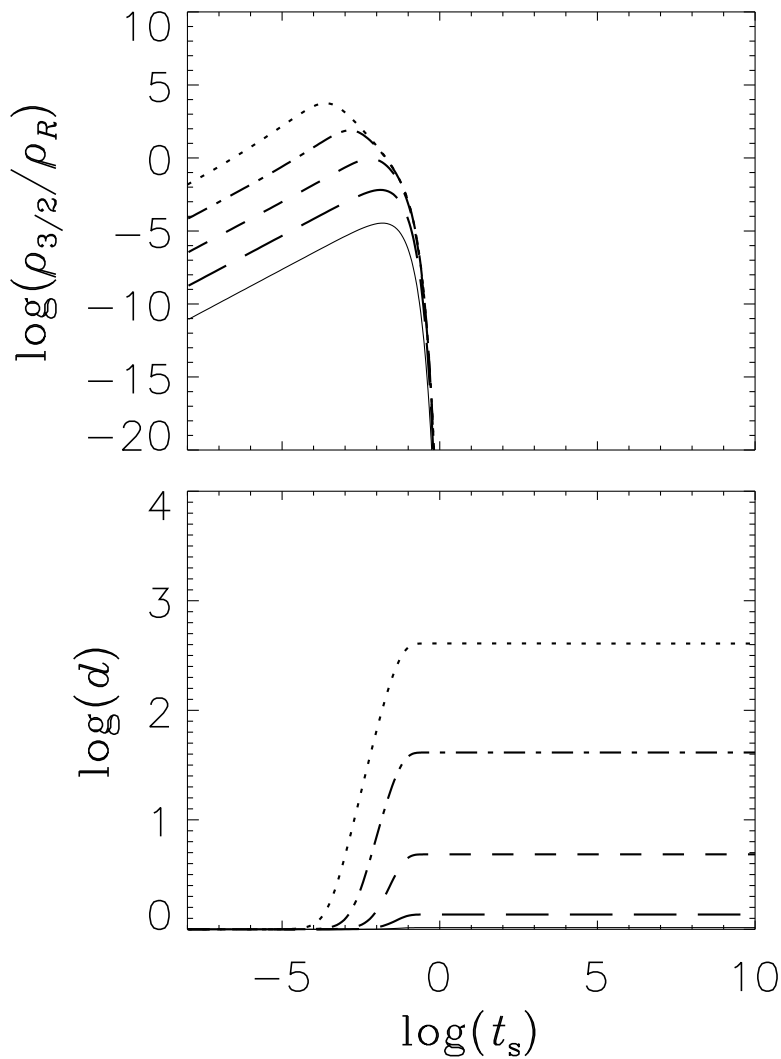}
\vspace{0.0truecm}
\end{center}
\caption{Plot of temperature, density and entropy increase as
functions of time for $m_{3/2}=100$ TeV. Curves are for
$(n_{3/2}/s)=10^{-10}, 10^{-9}, 10^{-8}, 10^{-7}$, and $10^{-6}$ in
increasing order.} \label{figure1}
\end{figure}

The entropy generated by gravitino decay can be estimated from
Eq.~(\ref{sfermdilution}). However, the evolution of gravitinos and
relativistic particles can be followed directly by solving the
Boltzmann equation. Assuming that gravitinos are non-relativistic at
decay and that $g_*$ is constant during the decay process, the
Boltzmann equations for gravitinos take the form
\begin{equation}
\dot{\rho}_{3/2} = -\Gamma_{3/2} \rho_{3/2} - 3 H \rho_{3/2}
\end{equation}
and
\begin{equation}
\dot{T} = -H T + \Gamma_{3/2} \frac{\rho_{3/2} T}{4 \rho_{\rm R}}.
\end{equation}
These should be solved together with the Friedmann equation
\begin{equation}
H^2 = \frac{1}{3 M^2_{\rm P}}\left[\rho_{3/2} + g_*(T) \frac{\pi^2}{30} 
T^4\right].
\end{equation}
In Fig.~(\ref{figure1}) we show several examples of solving the
Boltzmann equation for different initial conditions, always assuming
the $g_* = 10.75$ during decay.  The dilution factor $d$ can then be
found from
\begin{equation}
d(t) = \frac{(a(t)T(t))^3}{(a_{\rm i} T_{\rm i})^3},
\end{equation}
where $a_{\rm i}$ and $T_{\rm i}$ denote the initial values of the
scale factor and the radiation temperature respectively. The bottom
panel of Fig.~(\ref{figure1}) shows the dilution factor, and when
gravitinos dominate it agrees quite well with
Eq.~(\ref{sfermdilution}).

\section{Leptogenesis}

The baryon asymmetry of the Universe (BAU) parameterized as $\eta_{\rm
B}=(n_{\rm B}-n_{\bar{\rm B}})/s$, with $s$ being the entropy density,
is determined to be $0.9 \times 10^{-10}$ by recent analysis of WMAP
data~\cite{WMAP}. This number is also in good agreement with an
independent determination from primordial abundances produced during
BBN~\cite{cfo}. Three conditions are required for generating a baryon
asymmetry: $B-$ and/or $L-$violation, $C-$ and $CP-$violation, and
departure from thermal equilibrium~\cite{sakharov}.  Since
$B+L$-violating sphalerons transitions are active at temperatures
$100~{\rm GeV} \leq T \leq 10^{12}$ GeV~\cite{krs}, any mechanism for
creating a baryon asymmetry at $T > 100$ GeV must create a $B-L$
asymmetry. The final asymmetry is then given by $B=a(B-L)$, where
$a=28/79$ in case of SM and $a=8/23$ for MSSM~\cite{khlebnikov}.

Leptogenesis postulates the existence of RH neutrinos, which are SM
singlets, with a lepton number violating Majorana mass
$M_N$. It can be naturally embedded in models 
which explain the light neutrino masses
via the see-saw mechanism~\cite{seesaw}. A lepton asymmetry can then
be generated from the out-of-equilibrium decay of the RH neutrinos into
Higgs bosons and light leptons, provided $CP-$violating phases
exist in the neutrino Yukawa couplings~\cite{fy,luty,one-loop}. 
The created lepton asymmetry will be
converted into a baryonic asymmetry via sphalerons
processes.

The on-shell RH neutrinos whose decay is responsible for the lepton
asymmetry can be produced thermally
via their Yukawa interactions with the standard model fields and their
superpartners~\cite{plumacher}, for which $T_{\rm R}\geq M_{1}\sim
10^{9}$~GeV,~\cite{buchmuller,sacha,gnrrs,strumia}, or non-thermally
for which $T_{\rm R}\leq M_{N}$,
see~\cite{infl,reheat,gprt}. Non-thermal leptogenesis can also
be achieved without exciting on-shell RH neutrinos~\cite{off-shell}.
In supersymmetric models one in addition has the RH sneutrinos which
serve as an additional source for leptogenesis~\cite{cdo}. The
sneutrinos are produced along with neutrinos in a thermal bath or
during reheating, and with much higher abundances in
preheating~\cite{bdps}. There are also additional possibilities for
leptogenesis from the RH sneutrinos some of which rely on soft SUSY
breaking effects~\cite{my,hmy,bmp,adm,ad3,soft}.


\subsection{Thermal Leptogenesis}

Let us concentrate on the supersymmetric standard model augmented with
three RH neutrino multiplets in order to accommodate neutrino masses
via the see-saw mechanism~\cite{seesaw}. The relevant part of the
superpotential is
\beq \label{superpot}
W \supset {1 \over 2} M_i {\bf N}_i {\bf N}_i + {\bf h}_{ij} {\bf H}_u 
{\bf N}_i {\bf L}_j,
\eeq
where ${\bf N}$, ${\bf H}_u$, and ${\bf L}$ are multiplets containing
the RH neutrinos $N$ and sneutrinos $\tilde N$, the Higgs field giving
mass to, e.g., the top quark and its superpartner, and the left-handed
(s)lepton doublets, respectively. Here ${\bf h}_{ij}$ are the neutrino
Yukawa couplings and we work in the basis in which the Majorana mass
matrix is diagonal. The decay of a RH (s)neutrino with mass $M_i$ (we
choose $M_1 < M_2 < M_3$) results in a lepton asymmetry per
(s)neutrino quanta $\epsilon_i$, given by
\beq \label{asymmetry}
\epsilon_i = - {1 \over 8 \pi} {1 \over [{\bf h} {\bf h}^{\dagger}]_{ii}} 
\sum_{j} {\rm Im}
\left( [{\bf h} {\bf h}^{\dagger}]_{ij}\right)^2 f \left({M^{2}_{j}
\over M^{2}_{i}} \right)\,,
\eeq          
with~\cite{one-loop}
\beq \label{corrections}
f(x) = \sqrt{x} \left ({2 \over x-1} + {\ln} \left[{1 + x \over
x}\right ] \right ).
\eeq
The first and second terms on the right-hand side of
Eq.~(\ref{corrections}) correspond to the one-loop self-energy and
vertex corrections, respectively.  Assuming strongly hierarchical RH
(s)neutrinos and an ${\cal O}(1)$ $CP-$violating phase in the Yukawa
couplings, it can be shown that~\cite{di}
\beq \label{hierarchical}
\vert \epsilon_1 \vert \lsim {3 \over 8 \pi} {M_1 (m_3 - m_1) \over 
{\langle H_u 
\rangle}^{2}}\,,
\eeq
where $m_1 < m_2 < m_3$ are the masses of light, mostly left-handed
(LH) neutrinos. For a hierarchical spectrum of light neutrino masses
($m_1 \ll m_2 \ll m_3$),
we then have
\beq \label{epsilon1}
\vert \epsilon_1 \vert \lsim 2 \times 10^{-7} 
\left({m_3 - m_1 \over 0.05~{\rm eV}}\right) \left({M_1 \over 
10^{9}~{\rm GeV}}\right)\,.
\eeq
To obtain this, we have used $m_3 - m_1 \simeq m_3\simeq 0.05$ eV (as suggested
by atmospheric neutrino oscillation data) and $\langle H_u \rangle \simeq
170$ GeV.

If the asymmetry is mainly produced from the decay of the lightest RH
states, after taking the conversion by sphalerons into account, we
arrive at
\beq \label{baryonthermal}
\eta_{\rm B}^{\mbox{}_{\rm MAX}} \simeq 3 \times 10^{-10} 
\left({m_3 - m_1 \over 0.05~{\rm eV}}\right) 
\left({M_1 \over 10^9~{\rm GeV}}\right) \kappa \, ,
\eeq
where we have assumed maximal $CP$-violation. Here $\kappa$ is the
efficiency factor accounting for the decay, inverse decay and
scattering processes involving the RH states~\cite{buchmuller,gnrrs}.

The decay parameter $K$ is defined as
\beq \label{decpar}
K \equiv {\Gamma_1 \over H(T = M_1)},
\eeq
where 
\beq \label{Ndec}
\Gamma_1 = \sum_{i} {{\vert h_{1i} \vert}^2 \over 4 \pi} M_1,
\eeq
is the decay width of $N_1$ and ${\tilde N}_1$. One can also define
the effective neutrino mass
\beq \label{effmass}
{\tilde m}_1 \equiv \sum_{i} {{\vert h_{1i} \vert}^2 {\langle H_u \rangle}^2 
\over M_1},
\eeq
which determines the strength of ${\tilde N}_1$ and $N_1$
interactions, with the model-independent bound $m_1 < {\tilde
m}_1$~\cite{fhy}.

\begin{itemize}

\item{{\it case(1)}:\\ If $K < 1$, corresponding to ${\tilde m}_1 <
10^{-3}$ eV, the decay of RH states will be out-of-equilibrium at all
times. In this case the abundance of RH states produced via Yukawa
interactions does not reach the thermal equilibrium value.  The lepton
number violating scatterings can be safely neglected.
Hence this case is called
the weak washout regime. The efficiency factor is  $\kappa \simeq 0.1$
when ${\tilde m}_1 = 10^{-3}$ eV. Generating sufficient asymmetry then
puts an absolute lower bound $few \times 10^9$ GeV on $M_1$, and
$T_{\rm R} \geq M_1$ will be required in this case~\cite{buchmuller}.}

\item{{\it case (2)}:\\ In the opposite limit $K > 1$, ${\tilde N}_1$
and $N_1$ will be in thermal equilibrium at temperatures $T > M_1$. In
particular, the efficiency of inverse decays erases any pre-existing
asymmetry (generated, for example, from the decay of heavier RH
states). This regime is called of strong washout.  We note that this
regime includes the entire favored neutrino mass range $m_{\rm sol} ~
\lsim {\tilde m}_1 ~ \lsim m_{\rm atm}$. The inverse decays keep the
RH (s)neutrinos in equilibrium for sometime after $T$ drops below
$M_1$. The number density of quanta which undergo out-of-equilibrium
decay is therefore suppressed and reduces the efficiency factor. 
Successful leptogenesis in the range $(m_{\rm sol},m_{\rm atm})$
requires that $10^{10}~{\rm GeV} < M_1 \lsim 10^{11}$ GeV while, due
to the efficiency of inverse decays, $T_{\rm R}$ can be smaller than
$M_1$ by almost one order of magnitude~\cite{buchmuller}. The
efficiency factor $\kappa$ in this window varies between $few \times
10^{-3}$ and $few \times 10^{-2}$.}
\end{itemize}

It is possible to obtain (approximate) analytical expressions for the
efficiency factor $\kappa$. In the strong washout regime, the final
efficiency factor is maximal when $\Delta L = 2$ scatterings among the
LH (s)leptons can be neglected. So long as the scatterings can be
neglected, the efficiency factor $\kappa$ is independent from the lightest RH
(s)neutrino mass $M_1$ and 
$\kappa(\tilde{m}_1)$
is given by~\cite{buchmuller,gnrrs,pilaftsis,strumia}
\begin{equation}
\kappa \simeq  10^{-2} \left( \frac{0.01 \ {\rm eV}}{\tilde{m}_1} 
\right)^{1.1}\,.
\label{efficiency}
\end{equation}
Scatterings cannot be neglected for very large $M_1$ or large
light neutrino masses. In this case the efficiency factor can be
approximated as~\cite{buchmuller}:
\begin{equation}
\label{efficiencycomplete}
\kappa(\tilde{m}_1, M_1, \bar{m}^2) = \kappa (\tilde{m}_1) 
e^{- \frac{\omega}{z_{\rm B}}
\left( \frac{M_1}{10^{10} \ {\rm GeV} }\right)\left({{\bar m} \over 
1~{\rm eV}}\right)^2},
\end{equation}
where $\omega \simeq 0.186$, $z_{\rm B}= M_1/T_{\rm B}\sim $~few, with
$T_{\rm B}$ the temperature at which most of the lepton asymmetry is
produced, $\bar{m}^2$ is the sum over the squares of the light
neutrino masses.  The efficiency factor $\kappa (\tilde{m}_1) $ is given in
Eq.~(\ref{efficiency}). For hierarchical light neutrinos, $\bar{m}^2 = 
m_3^2 \simeq
\Delta m^2_{atm} \simeq 2.2 \times 10^{-3} \ {\rm eV}^2$,
where $\Delta m^2_{atm}$ is the mass squared difference 
which controls the oscillations of atmospheric neutrinos. Washout
effects due to scatterings then become important for $M_1 \sim
10^{14}$~GeV. For quasi-degenerate neutrinos, $m_1 \simeq m_2 \simeq
m_3$, larger values of ${\bar m}$ are possible and $\kappa$ will be 
exponentially
suppressed even for much smaller values of $M_1$.

For negligible $\Delta L = 2$ scatterings,
the baryon asymmetry is, in the case of maximal decay asymmetry,
given by
\beq \label{baryonlargeM1}
{\eta_{\rm B}^{\mbox{}_{\rm MAX}}} \simeq 0.9 \times 10^{-10} \ \left({m_3 - 
m_1 \over 0.05~ 
{\rm eV}}\right) \left( \frac{0.01 \ {\rm eV}}{\tilde{m}_1} \right)^{1.1} 
\left({M_1 \over 3.7 \times 10^{10}~{\rm GeV}}\right).
\end{equation}
The baryon asymmetry is proportional to $M_1$. The usual bound on the
reheating temperature $T_{\rm R} \leq 10^{10}$ GeV, given for $m_{3/2}
< 50$ TeV, does not apply to supermassive gravitinos. Hence, RH
(s)neutrinos with masses $M_1 \gg 10^{10}$~GeV can be thermally produced in the
early Universe. This can lead to larger amounts of baryon asymmetry
generated by thermal leptogenesis, with respect to the standard
scenario. The lowest reheating temperatures required for thermally
producing these RH (s)neutrinos is $T_R \sim M_1/{few} \ll 10^{14}
$~GeV. Note that the initial assumption of gravitino non-domination is
satisfied if gravitinos are dominantly produced by thermal
scatterings.


\subsection{Effects of the Gravitino on Thermal Leptogenesis}

Thermal leptogenesis completes when $T \sim M_1/
{few}$~\cite{buchmuller,gnrrs}. Eq.~(\ref{dectemp}) then implies that
gravitino decay takes place after leptogenesis unless they are
extremely heavy:
\begin{equation} \label{before}
m_{3/2} > \left( \frac{M_1}{10^9 \ {\rm GeV}} 
\right)^{2/3} 10^{12} \ {\rm GeV}.
\end{equation}
On the other hand, for $m_{3/2} \geq 50$ TeV, gravitino decay occurs
at a temperature $T_{3/2} > 6.8 $ MeV which is compatible with a
successful BBN. Therefore both scenarios of gravitino domination and
non-domination are in agreement with the BBN constraints. Nevertheless
the effect of gravitino decay on the final baryon asymmetry need to be
taken into account. We consider both scenarios of gravitino domination
and non-domination.


\begin{itemize}

\item{{\it Gravitino non-domination}:\\ 
The condition for gravitino non-domination is given in
Eq.~(\ref{nodom}).  There will be practically no dilution by gravitino
decay in this case, and thermal leptogenesis should generate
$\eta_{\rm B} \simeq 10^{-10}$ according to
Eq.~(\ref{baryonthermal}). Obtaining sufficient asymmetry in both of
the weak and strong washout regimes requires that $M_1 \gsim T_{\rm R}
> 10^9$ GeV, and leptogenesis completes when $T \sim M_1 \geq 10^9$
GeV~\cite{buchmuller,gnrrs}. 
Sfermions with a
mass ${\tilde m} \leq 10^9$ GeV certainly reach thermal equilibrium
and, for $m_{3/2} < {\tilde m}$, their decay will produce gravitinos
according to Eq.~(\ref{decay}).

If $m_{3/2} \geq {\tilde m}$, gravitino production from sfermion
decays is kinematically forbidden. Scatterings of gauge and gaugino
quanta in the thermal bath will nevertheless produce gravitinos so
long as $m_{3/2} \leq T_{\rm R}$. Late time domination of gravitinos
thus produced requires that the reheating temperature $T_{\rm R} \geq
10^{14}$ GeV, see Eq.~(\ref{scattering}). However, gravitinos can be
overproduced for much smaller $T_{\rm R}$. For $m_{\chi} = 100~{\rm
GeV}~(1$ TeV), gravitino decay results in LSP overproduction when
$T_{\rm R} \geq 3 \times 10^{10}~(3 \times 10^{9})$ GeV.  This indeed
occurs for the bulk of the parameter space compatible with thermal
leptogenesis, particularly in the favored neutrino mass window
$m_{\rm solar} \leq {\tilde m}_1 \leq m_{\rm atm}$~\cite{buchmuller}.
The condition for sufficient annihilation of non-thermal LSPs in this
case sets a lower bound on the gravitino mass through Eq.~(\ref{eff}),
independently of whether $m_{3/2} < {\tilde m}$ or $m_{3/2} \geq
{\tilde m}$. Constraints from LSP annihilation, which determine acceptable 
regions of the ${\widetilde m} - m_{3/2}$ plane, are summarized in Fig.~(1) and
the related discussion.

In the case of gravitino non-domination, the produced
lepton asymmetry is not subsequently 
diluted by gravitino decays, even if taking place after leptogenesis.
As the bound on the reheating temperature $T_{\rm R} \leq 10^{10}$~GeV
does not apply for supermassive gravitinos,
RH neutrinos with masses $M_1 \gg 10^{10}$~GeV can be fully thermalized 
for a sufficiently large reheating temperature.
From Eqs.~(\ref{baryonthermal}), (\ref{efficiency}) and 
(\ref{efficiencycomplete}),
it follows that the baryon asymmetry is proportional
to the lightest RH neutrino mass, $M_1$,
up to $M_1 \sim 10^{14}$~GeV,
for $\bar{m}^2 \sim \Delta m^2_{atm}$.
For larger values of $M_1$,
the lepton asymmetry is washed out by
$\Delta L=2$ scatterings.
Then, the maximum baryon asymmetry is produced for
$M_1 \sim 5 \times 10^{12} \ z_{\rm B} /\omega$~GeV.
Here we have again taken $\bar{m}^2 \sim \Delta m^2_{atm}$.
From Eq.~(\ref{baryonlargeM1}), we notice that for large values of $M_1$
the generated $\eta_{\rm B}$ can be much larger
than the one required to explain the observations,
if the decay asymmetry is maximal.
Models which implement 
the see-saw mechanism of neutrino mass generation
typically
assume the conservation of flavor symmetries
and/or special forms of the Yukawa couplings,
in order to explain the low energy
neutrino masses and mixing.
In many of these models,
the decay asymmetry is constrained to be non maximal
and $M_1$ larger than the typical values $10^9$--$10^{10}$~GeV 
are needed to generate a sufficient baryon asymmetry.
Models with supermassive gravitinos allow to have reheating temperatures
high enough to thermally produce such heavy RH neutrinos.
In each specific model,
a detailed analysis is required
for establishing the feasibility of successfull leptogenesis
and, at the same time, 
the possibility to explain the low energy neutrino mass matrix
(for a discussion of $CP$-violation in specific see-saw models and 
leptogenesis, see, e.g., Ref.~\cite{CPVconn1}).
}

\item{{\it Gravitino-dominated Universe}:\\   
If gravitinos are produced very abundantly, see Eq.~(\ref{dom}), the
Universe will become gravitino-dominated.  Sfermion decays (which, as
mentioned earlier, usually dominate over thermal scatterings and
inflaton decay) produce such abundances of gravitinos
when Eq.~(\ref{sfermdom}) is satisfied.

Gravitino decay reheats the Universe to a temperature $T_{3/2}$,
see Eq.~(\ref{dectemp}), and increases the entropy density by a factor $d$
given in Eq.~(\ref{dilution}). Successful thermal leptogenesis {\it after}
gravitino decay will be only possible if gravitinos are extremely
heavy $m_{3/2} > 10^{12}$ GeV, see Eq.~(\ref{before}). According
to~Eq.~({\ref{dom}), gravitino domination in this case requires that
$\left(n_{3/2}/s \right) > 10^{-5}$. It follows
from Eq.~(\ref{sfermdom}) that sfermion decays yield this only if ${\tilde
m}$ (and $T_{\rm R}$) is $> 10^{14}$~GeV. In addition,
non-perturbative production of gravitinos with the necessary abundance
is also questionable.

Therefore, it is more realistic to consider the opposite situation
where leptogenesis occurs {\it before} gravitino decay. In this case,
due to the entropy release by gravitino decay, the generated asymmetry
must exceed the observed value of $\eta_{\rm B} \simeq 10^{-10}$ by a
factor of $d$. Eqs.~(\ref{sfermdilution}) and~(\ref{baryonthermal})
then imply that the final asymmetry is
\beq \label{domasym}
\eta_{\rm B} \simeq 3 \times 10^{-10} 
\left({m_{3/2} \over 10^5~{\rm GeV}}\right)^{5/2} 
\left({1.3 \times 10^8~{\rm GeV} \over {\widetilde m}}\right)^3 
\left({M_1 \over 10^9~{\rm GeV}}\right) \kappa.
\eeq
As a specific example, consider leptogenesis in the favored neutrino
mass window $m_{\rm sol} \leq {\tilde m}_1 \leq m_{\rm atm}$. In this
interval, which entirely lies within the strong washout regime, the
efficiency factor $\kappa \sim 10^{-2}$, and the reheating temperature
follows $T_{\rm R} \geq 0.1 M_1$~\cite{buchmuller}. Therefore
successful leptogenesis requires that
\beq \label{suclep}
M_1 \simeq \left({10^5~{\rm GeV} \over m_{3/2}}\right)^{5/2} 
\left({{\widetilde m} \over 1.3 \times 10^8~{\rm GeV}}\right)^3
3 \times 10^{10}~{\rm GeV}.
\eeq

In Fig.~\ref{figure3} we show the value of $M_1$ needed to produce the
correct baryon asymmetry of $\eta_{\rm B} = 0.9 \times 10^{-10}$ as a
function of the lightest RH (s)neutrino mass $m_{3/2}$ and
${\widetilde m}$.  The plot is produced assuming $\kappa = 10^{-2}$,
but since $\eta_{\rm B} \propto \kappa$ it can easily be rescaled for
other values. Without dilution, successful leptogenesis for $\kappa =
10^{-2}$ requires that $M_1 \simeq 3 \times 10^{10}$ GeV, see
Eq.~(\ref{baryonthermal}). Therefore having contours with $M_1 > 3
\times 10^{10}$ GeV indicates gravitino domination. As mentioned
earlier, thermal leptogenesis fails for $M_1 \geq 10^{14}$ GeV due to
the erasure of generated asymmetry by $\Delta L = 2$ scattering
processes. This happens in the light colored region, and hence excludes it. 
The overlap between Figs.~(1) and~(3)
combines the constraints from leptogenesis and dark matter
considerations.

Thermal leptogenesis cannot generate a baryon asymmetry which exceeds
$10^{-2}$. The maximal value is obtained in the (hypothetical) case
when RH (s)neutrinos have thermal equilibrium abundance, and the
efficiency factor $\kappa$ and asymmetry parameter $\vert \epsilon_{1}
\vert$ are both $1$, see Eq.~(\ref{baryonthermal}). This implies that
successful leptogenesis in a gravitino-dominated Universe would be
impossible if the dilution factor $d$ was larger than $10^8$. However,
since gravitinos can at most reach thermal equilibrium, we always have
$d \leq 5.9 \times 10^5$, see the discussion after
Eq.~(\ref{sfermdilution}). This ensures that there will be no such
case where thermal leptogenesis is absolutely impossible in a
gravitino-dominated Universe.

}
}
\end{itemize}


\begin{figure}[htb]
\vspace*{-0.0cm}
\begin{center}
\epsfysize=9truecm\epsfbox{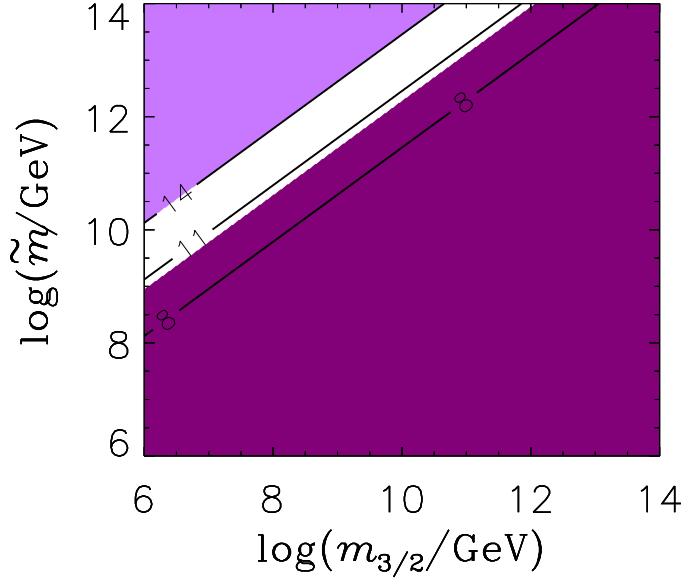}
\vspace{0.0truecm}
\end{center}
\caption{The value of $\log(M_1)$ (in GeV) needed to produce
$\eta_{\rm B} = 0.9 \times 10^{-10}$ as a function of $m_{3/2}$ and
$\widetilde{m}$ in the gravitino-dominated scenario. We have chosen
the efficiency factor $\kappa = 10^{-2}$, which is the typical value in
the favored neutrino mass window. The dark color region does not have
gravitino domination, while in the light color region thermal 
lepotgenesis fails due to washout by lepton number violating scatterings.}
\label{figure3}
\end{figure}


\section{Baryogenesis from Supersymmetric Flat Directions}

There are many gauge invariant combinations of the Higgs, squark and
slepton fields along which the scalar potential identically vanishes
in the limit of exact SUSY. Within the MSSM there are nearly $300$
flat directions which are both $F-$ and $D-$ flat and conserve
R-parity~\cite{gkm}. Soft terms, as well as non-renormalizable
superpotential terms, lift the flat directions when SUSY is broken.

A homogeneous condensate along can be formed along a flat direction in
the inflationary epoch, provided that the flat direction mass
${\widetilde m}$ is smaller than the Hubble expansion rate during
inflation.  The condensate starts oscillating coherently when the
expansion rate $H \simeq {\widetilde m}$. During this epoch the
inflationary fluctuations of the condensate can be converted to
density perturbations~\cite{denspert}. The condensate can also help an
efficient reheating to the SM degrees of freedom~\cite{averdi}. In
addition, it can excite vector perturbations to explain the observed
large scale magnetic field~\cite{mag}. If the condensate carries a
non-zero baryon and/or lepton number, then the flat direction dynamics
can be responsible for baryogenesis via the Affleck-Dine (AD)
mechanism~\cite{drt} (for a review, see Ref.~\cite{Kari}). In the
following section we make a general discussion on the viability of AD
baryogenesis for arbitrarily heavy gravitinos and/or sfermions.


\subsection{Late Baryogenesis via the Affleck--Dine Mechanism}

The scalar potential for a flat direction $\phi$ (not to be confused
with the inflaton field in Section II) is given by~\cite{drt}
\beq
\label{adpot}
V = \left(\widetilde m^2 + c_H H^2 \right)|\phi|^2 + 
\left(\frac{ A + a H}{n M^{n-3}}\lambda \phi^n + h.c. \right)
+\frac{\lambda^2}{M^{(2n-3)}}|\phi|^{2(n-1)}\,.
\eeq
Here ${\widetilde m}^2$ and $c_H H^2$ are the soft ${\rm mass}^2$ from
SUSY breaking in the vacuum and SUSY breaking by the non-zero energy
density of the inflaton respectively. Here $c_H$ can have either sign,
and its sign is also affected by radiative
corrections~\cite{Gaillard,manuel}. The last term on the right-hand
side of~(\ref{adpot}), which lifts the flat direction, arises from a
non-renormalizable superpotential term (i.e. $n \geq 4$) induced by
new physics at a high scale $M$. In general $M$ could be a string
scale, below which we can trust the effective field theory, or $M =
M_{\rm P}$.  SUSY breaking in the vacuum and by the inflaton energy
density generate $A$-terms $A$ and $a H$, respectively, corresponding
to this non-renormalizable superpotential term.  For minimal K\"ahler
terms and in case of gravity mediation $ {\widetilde m} \sim A \sim
m_{3/2}$ and $0 < c_H \sim 1$. Depending on the symmetries of the
inflaton sector, both $a \sim {\cal O}(1)$ and $a \ll 1$ are possible.

The equation of motion for the flat direction is given by
\beq
\label{adeqm}
\ddot\phi + 3 H \dot \phi + \frac{\partial V}{\partial\phi^*} = 0\,.
\eeq
The evolution is easiest to analyze by the field parameterization
\beq
\label{param}
\phi = \frac{1}{\sqrt{2}}\,\varphi e^{i\theta}\,,
\eeq
where $\varphi,\,\theta$ are real fields. Then the scalar potential
can be written in the form
\beq
\label{adpot2}
V(\varphi,\theta) = \frac{1}{2} \left( \widetilde m^2 + c_H H^2 \right)
\varphi^2 + \frac{|\lambda| f(\theta) }{2^{(n-2)/2} n M^{n-3}} \,
\varphi^n + \frac{|\lambda|^2}{2^{n-1} M^{2(n-3)}} \,
\varphi^{2(n-1)}\,. 
\eeq
Here
\beq
\label{ftheta}
f(\theta) = |A| \cos(n\theta + \theta_A + \theta_{\lambda} ) + |a| H
\cos(n \theta + \theta_a + \theta_{\lambda})\,,
\eeq
with $\theta_A,~\theta_a,~\theta_{\lambda}$ being the angular
directions for $A,~a,~\lambda$ respectively. The baryon/lepton
number density is given by
\beq
\label{charge}
n_{\rm B,L} = \frac{\beta}{i} \left( \phi^* \dot\phi - \phi\dot\phi^*
\right) = \beta \dot\theta \, \varphi^2\,, 
\eeq
where $\beta$ is the baryon/lepton number carried by the flat
direction. At the minima of the of potential
\beq
\label{extrep}
\varphi_{\rm min}^{n-2} = \frac{2^{n/n-2}\,M^{n-3}}{(n-1)\,|\lambda|}
\left\{ -f(\theta) \pm \left[f(\theta)^2 - 4(n-1)(\widetilde m^2 + c_H H^2)
\right]^{1/2} \right\}\,, 
\eeq
and $n \theta_{\rm min} = (2p+1)\pi - \theta_a -\theta_{\lambda}$ if
$|a|H \gg |A|$, while $n \theta_{\rm min} = (2p+1)\pi - \theta_A -
\theta_{\lambda}$ if $|a|H \ll |A|$ (with $p=0,1,\ldots,n-1$).

The radial field $\varphi$ quickly settles at one of the minima given
in~(\ref{extrep}) during inflation. If $|a|H \gg |A|$, the phase field
$\varphi \theta$ (since $\theta$ is dimensionless) has a mass of order
$|a|H$ and it ends up in one of the discrete minima $n \theta_{\min} =
\pi -\theta_a -\theta_{\lambda}$. The phase field has a mass $\ll H$
if $|a| H \ll |A|$, and hence it freezes at a random value.

After inflation the Hubble rate decreases as the Universe expands, and
so does $\varphi_{\rm min}$. The $\varphi$ field tracks the
instantaneous minimum of the potential, so its evolution can be
qualitatively understood by looking at the evolution of the
minimum. Once ${\widetilde m}^2 \simeq |c_H| H^2$, the minimum of the
potential changes from $\varphi_{\rm min}$ to $\varphi = 0$ in a
non-adiabatic manner. At this time $\phi$ starts oscillating in the
radial direction with frequency ${\widetilde m}$.  The motion of the
phase field, which is necessary for generating a baryon/lepton
asymmetry, requires the exertion of a torque. If $|a|H \sim |A|$, a
non-adiabatic change in the position of the minimum from $n
\theta_{\rm min} = \pi - \theta_a - \theta_{\lambda}$ to $n
\theta_{\rm min} = \pi - \theta_A -\theta_{\lambda}$ generates a
torque.  If $|a| H \ll |A|$, the freezing of the phase field at a
random value generates the torque and leads to its motion towards the
minimum $n \theta_{\rm min} = \pi - \theta_A -\theta_{\lambda}$ if
$|a| H \ll |A|$. The potential along the angular direction will
quickly decrease due to the redshift of $\varphi$, see
Eq.~(\ref{adpot2}), once $\varphi$ starts its oscillations. In
consequence, $\phi$ starts freely rotating in the angular direction at
which time a net baryon/lepton asymmetry is generated.

Based on Eqs.~(\ref{adeqm}) and~(\ref{charge}), the baryon/lepton
asymmetry obeys the equation
\beq
\label{cheqm}
\dot n_{\rm B,L} + 3H n_{\rm B,L} = -\beta \frac{\partial V}{\partial\theta}\,.
\eeq
This can be integrated to give at late times $t \gg H^{-1}_{\rm osc}$  
(see~\cite{Kari})
\beq
\label{chdens}
n_{\rm B,L} \simeq \beta \frac{2(n-2)}{3(n-3)} {\sin \delta \over 
(H_{\rm osc} t)^2}|A| \varphi_{\rm osc}^2 \,,
\eeq
where $\varphi_{\rm osc},\, H_{\rm osc}$ denote the value of $\varphi,
\, H$ when the condensate starts oscillating. Here $\delta$ is a
measure of spontaneous $CP$-violation in the $\phi$ potential and
$\sin \delta \sim 1$.  The baryon to entropy ratio will then be
\beq
\label{barentr}
\frac{n_{\rm B,L}}{s} = \frac{3 T_{\rm R} \, n_{\rm B,L}}{4\rho_{\rm R}} = 
\frac{T_{\rm R} \, n_{\rm B,L}}{4 M_{\rm P}^2 H_{\rm osc}^2}\,. 
\eeq
We parameterize the $A$-term as $|A| = \gamma \widetilde m$. For
gravity-mediated SUSY breaking typically ${\widetilde m} \sim |A| \sim
m_{3/2}$, while in gauge-mediated models $|A| \sim m_{3/2} \ll
\widetilde m$. In split SUSY $A \ll \widetilde m$, see the discussion
in Ref.~\cite{adgr}. Here we consider $\gamma$ to be arbitrary, and
hence $\widetilde m$ and $|A|$ be unrelated.

Since $H_{\rm osc} \simeq {\widetilde m}$ and $\varphi_{\rm
osc}^{n-2}\sim 2^{(n-2)/2} M^{n-3} \widetilde m / |\lambda|$, we find
\beq
\label{barentr2}
\frac{n_{\rm B,L}}{s} \simeq \frac{n-2}{6(n-3)} \,
\frac{|A|T_{\rm R}}{\widetilde m^2} \left( \frac{\widetilde m}{|\lambda|
M_{\rm P}} \right)^{2/(n-2)}\,, 
\eeq
where $M = M_{\rm P}$ is taken.

For the lowest dimensional non-renormalizable term $n=4$, and
$\lambda\sim {\cal O}(1)$, the generated baryon asymmetry is $n_{\rm
B,L}/s\sim 0.1(|A|T_{\rm R}/\widetilde m M_{\rm P})$. If $|A| \sim
\widetilde m$ then $T_{\rm R} \sim 10^{9}$~GeV is adequate to generate
baryon asymmetry of order $10^{-10}$. If $|A| \ll \widetilde m$, one
would require even larger $T_{\rm R}$.  However the most desirable
feature of AD baryogenesis lies in its flexibility to generate a
desirable baryon asymmetry even for very low reheating temperatures. For
instance, if $\widetilde m \sim 10^{7}$~GeV and $n=6$, the required
asymmetry can be generated for $T_{\rm R} \sim 10^{3}$~GeV when $|A|=
\widetilde m$.

At late times, i.e. $H \ll {\widetilde m}$, contributions from SUSY
breaking by the inflaton energy density are subdominant to soft terms
from SUSY breaking in the vacuum. If $|A|^2< 4(n-1) {\widetilde m}^2$,
the $\phi$ potential has only one minimum at $\varphi = 0$. However,
another minimum appears away from the origin when $4(n-1) {\widetilde
m}^2 \leq |A|^2$, see~(\ref{extrep}). In the AD scenario the $\phi$
field starts at large values of $\varphi$, and hence it gets trapped
in this secondary minimum in the course of its evolution in the early
Universe. If $4(n-1){\widetilde m}^2 \leq |A|^2 < n^2{\widetilde
m}^2$, the true minimum is still located at the origin.  Tunneling
from the false vacuum could still save the situation in this case.
However, for $|A| \geq n {\widetilde m}$, the true minimum will be at
$\varphi \neq 0$. Since flat directions have non-zero charge and
color quantum numbers, this will lead to an unacceptable situation
with charge and color breaking in the vacuum. It is therefore
necessary to have $|A|^2 < 4 (n-1) {\widetilde m}^2$ in order to avoid
entrapment in such vacuum states.  This is the case in
gravity-mediated and gauge-mediated models, as well as split
SUSY. Note that the same discussion applies to the soft terms induced
by non-zero energy density of the inflaton. However, these terms
disappear at late times and will be irrelevant in the present vacuum.
               

\begin{figure}[htb]
\vspace*{-0.0cm}
\begin{center}
\epsfysize=9truecm\epsfbox{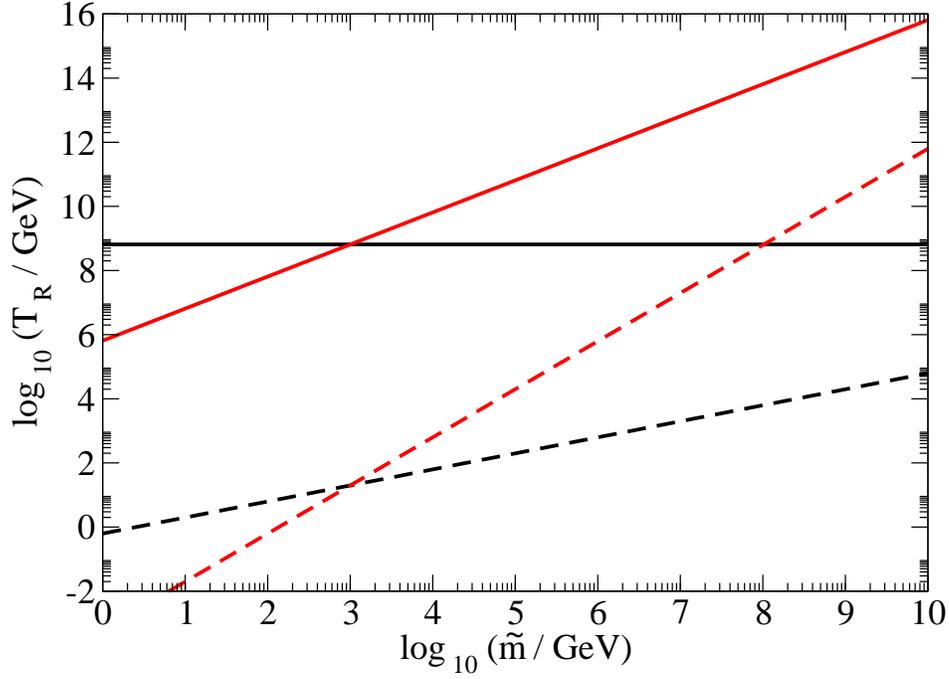}
\vspace{0.0truecm}
\end{center}
\caption{The reheating temperature as a function of the scalar mass,
given by Eq.~(\protect\ref{barentr2}), for successful AD baryogenesis. The
cases $n=4$ and $n=6$ are represented by solid and dashed lines
respectively. The black and red lines are plotted for $|A|= \widetilde
m$ and $|A|=1$~TeV respectively.}
\end{figure}

\begin{itemize}

\item{{\it Thermal effects}:\\ According to the potential given in
Eqs.~(\ref{adpot}) and~(\ref{adpot2}), the flat direction condensate
starts oscillating when $H \simeq {\widetilde m}$. However thermal
effects from reheating may trigger an earlier oscillation and lead to
a larger value of $H_{\rm osc}$~\cite{ace,and}. The inflaton decay (in
the perturbative regime) is a gradual process which starts after the
end of inflation. Hence, even before the inflaton decay is completed,
the decay products constitute a thermal bath with instantaneous
temperature $T \simeq (H T^2_{\rm R} M_{\rm P})^{1/4} > T_{\rm
R}$~\cite{kt}. The flat direction has gauge and Yukawa couplings,
collectively denoted by $y$, to other fields. Its VEV gives a mass
$\sim y \varphi$ to these fields. If $y \varphi \leq T$, these fields
are excited in the thermal bath and reach thermal equilibrium. This,
in turn, results in a thermal correction $\sim y T$ to the $\phi$
mass. If $y T$ exceeds the Hubble parameter at early times, i.e. for
$H \gg {\widetilde m}$, the condensate starts early
oscillations~\cite{ace}. On the other hand, if $y \varphi > T$, the
fields coupled to $\phi$ will be too heavy to be excited. They will
decouple from the running of gauge coupling(s) at temperature $T$
instead, which induces a logarithmic correction to the free energy
$\sim T^4 \log \left(T/\varphi \right)$. This triggers early
oscillations of the condensate if $T^2/\varphi > H$ when $H \gg
{\widetilde m}$~\cite{and}. Note that according to Eq.~(\ref{barentr})
a larger value of $H_{\rm osc}$ results in a smaller baryon/lepton
asymmetry.

Refs.~\cite{ace,and} have studied thermal effects for the conventional
case with ${\widetilde m} \sim 1$ TeV. Thermal corrections of the
former type will become less important as ${\widetilde m}$
increases. The reason is that $\varphi^{n-2} \propto H$,
see~(\ref{extrep}), and hence $y \varphi > H$ will be more difficult
to satisfy for larger ${\widetilde m}$.  Also, since $H \propto T^4$
at early times, $\left(T^2/\varphi \right)$ increases more slowly that
$H$. Therefore thermal corrections of the latter type will also become
less important when ${\widetilde m} \gg 1$ TeV.}

\end{itemize}


\subsection{Effects of Gravitino on Affleck-Dine Baryogenesis}

In Eq.~(\ref{barentr2}) no specific assumption is made about the
source which reheats the Universe. It can be either the inflaton
decay, as usually considered, or the decay of the flat direction
condensate. In case the inflaton decay reheats the Universe $T_{\rm R}
< {\widetilde m}$ and ${\widetilde m} \leq T_{\rm R}$ are both
possible. However, $T_{\rm R} \geq {\widetilde m}$ if the flat
direction is responsible for reheating the Universe. The energy
density in the condensate oscillations is ${\widetilde m}^2
{\varphi}^2$. The flat direction has gauge and Yukawa couplings to
other fields through which it induces a mass $\propto \varphi$ for the
decay products. The condensate will decay no later than the time when
$\varphi \lsim {\widetilde m}$. Hence, since energy density in the
radiation is $\sim T^4$, we will have $T_{\rm R} \geq {\widetilde m}$
in this case. According to Eq.~(\ref{barentr2}), the yielded baryon
asymmetry is $\propto T_{\rm R}$. This implies a larger asymmetry for
larger values of $T_{\rm R}$. Having an unacceptably large baryon
asymmetry is indeed typical when the flat direction condensate has a
very large VEV such that it dominates the energy density and,
subsequently, reheats the Universe~\cite{ad}. Gravitino domination
can in this case help to dilute the excess of baryon asymmetry. It is
interesting that such a solution, invoked from the early days of AD
baryogenesis~\cite{ad}, can be naturally realized with supermassive
gravitinos.

When $T_{\rm R} \geq {\widetilde m}$ sfermions reach thermal
equilibrium after reheating and their decay will be the dominant
source of gravitino production. The condition for gravitino domination
and the dilution factor from gravitino decay are then given by
Eqs.~(\ref{sfermdom}) and~(\ref{sfermdilution}) respectively. The
final asymmetry generated via the AD mechanism in case of gravitino
domination will then be, see~(\ref{barentr2})
\beq 
\label{barentr3}
\frac{n_{\rm B}}{s}\approx \frac{n-2}{6(n-3)}
\left(\frac{\widetilde m}{|\lambda| M_{\rm P}}\right)^{2/n-2}
\left(\frac{|A|}{\widetilde m}\right)
\left({m_{3/2} \over 10^5~{\rm GeV}}\right)^{5/2}
\left({1.3 \times 10^8~{\rm GeV} \over {\widetilde m}}\right)^3\,.
\eeq
The results for the two extreme cases $|A| = {\widetilde m}$ and $|A|
= 1$ TeV are summarized in Fig.~(6) when $n = 4,6$. The combined
constraints from baryogenesis and dark matter considerations will be
included in the overlap of Figs.~(1) and~(6).

If $T_{\rm R} \ll {\widetilde m}$, sfermion quanta will not be excited
in the thermal bath. However, inflaton decay can in this case result
in efficient production of gravitinos according to
Eq.~(\ref{phidecay}). One can then repeat the same steps to find an
expression for the initial asymmetry similar to that
in~(\ref{barentr3}). Such an expression, and plots similar to those in
Fig.~(6), will however depend on $\Delta m_{\phi}$ as well as
${\widetilde m}$ and $m_{3/2}$. This leads to a more complicated and
model-dependent situation. Moreover, the scenario with gravitino
domination will be more constrained when $T_{\rm R} \ll {\widetilde
m}$ (specially if $|A| \ll {\widetilde m}$). The initial baryon
asymmetry is already suppressed in this case, see~(\ref{barentr2}),
and gravitino decay may dilute it to unacceptably small values.


\begin{figure}[htb]
\vspace*{-0.0cm}
\begin{center}
\epsfysize=9truecm\epsfbox{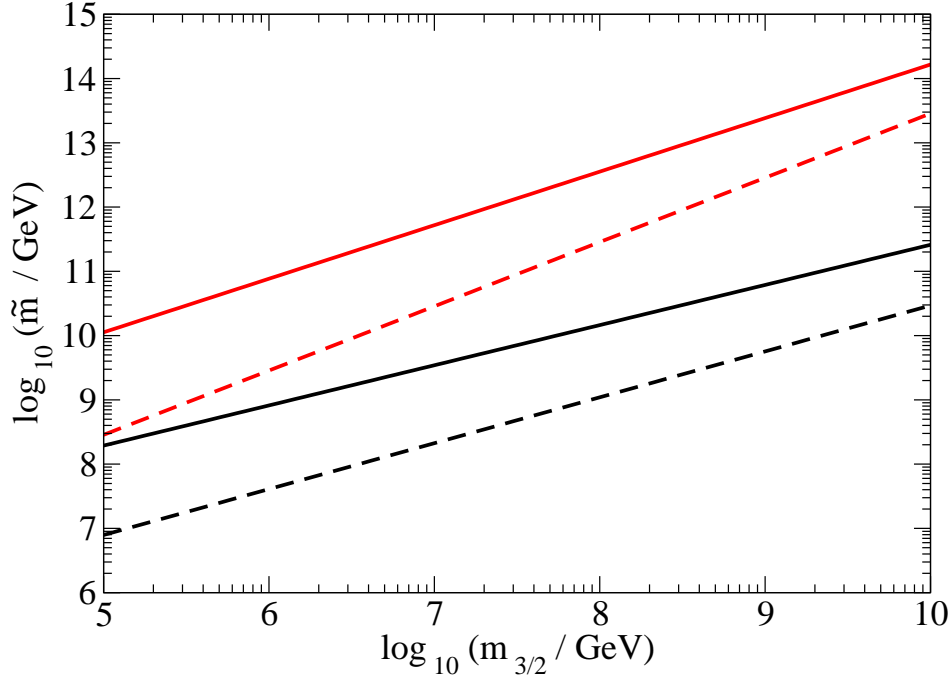}
\vspace{0.0truecm}
\end{center}
\caption{The scalar mass as a function of gravitino mass, given by 
Eq.~(\protect\ref{barentr3}), for successful AD 
baryogenesis in a gravitino-dominated Universe. The conventions are the same 
as in Fig.~(5).}
\end{figure}


\subsection{Late Gravitino Production from Q-ball Decay}

So far we have assumed that gravitinos are mainly produced in sfermion
decays (if $T_{\rm R} \geq {\widetilde m}$), or in inflaton decay (if
$T_{\rm R} \ll {\widetilde m}$). The flat direction condensate
consists of zero-mode quanta of the sfermions, and hence its decay too
can lead to gravitino production. If $T_{\rm R} \geq {\widetilde m}$,
this will be subdominant to the contribution from the decay of thermal
sfermions. The reason is that the zero-mode quanta have at most an
abundance $\left(n/s \right)$ which is comparable to that of thermal
sfermions. If $T_{\rm R} \ll {\widetilde m}$, sfermions will not be
excited in the thermal bath. As mentioned before, the condensate
certainly decays no later than the time when $\varphi \lsim
{\widetilde m}$. Even if $\varphi \sim M_{\rm P}$ initially, the
condition $\varphi \lsim {\widetilde m}$ is satisfied at $H \gsim
{\widetilde m}^2/M_{\rm P}$. Not that the scale factor of the Universe
$a \propto H^{-2/3}$ during reheating, in which phase the Universe is
dominated by inflaton oscillations. Also, the abundance of zero-mode
quanta $\left(n/s \right) < \left(3 T_{\rm R} /4 {\widetilde m}\right)
\ll 1$. This implies that the condensate contains a much smaller
number of quanta which survive (much) shorter than thermal
sfermions. Therefore gravitino production from the decay of the flat
direction condensate will not be as constraining as that in the decay
of thermal sfermion. In most case, it can be simply neglected.

The above discussion strictly applies to the oscillations of a
homogeneous condensate. However, it usually happens that the flat
direction oscillations fragment and forms non-topological solitons
known as Q-balls~\cite{coleman}.  These Q-balls can decay much later
than a homogeneous condensate. Their late decay could then efficiently
produce gravitinos, even if Q-balls do not dominate the energy density
of the Universe.

To elucidate, let us consider the case when tree-level sfermion masses
at a high scale $M$ are given by ${\widetilde m}$. The potential for
the sfermions, after taking into account of one-loop corrections,
reads
\beq
\label{gravqball}
V = \widetilde m^2 |\phi|^2 \left[ 1 + K \textrm{ln} 
\left( \frac{|\phi|^2}{M^2} \right) \right]\,,
\eeq
where $K$ is a coefficient determined by the renormalization group
equations, see~\cite{Nilles,enqvistetal}. In order to form Q-balls it
is necessary that the potential be flatter than $|\phi|^2$ at large
field values, i.e. that $K < 0$.  Loops which contain gauginos make a
negative contribution $\propto - m^2_{1/2}$ to $K$, with $m_{1/2}$
being the gaugino mass. On the other hand, loops which contain
sfermions contribute $\propto + {\widetilde m}^2$.  Then $K < 0$ can
be obtained, provided that $2 m_{1/2} \gsim {\widetilde
m}$~\cite{enqvistetal}. In models of gravity-mediated SUSY breaking
$\widetilde m \sim m_{1/2}$, and hence $K<0$ is obtained for many flat
directions.  When the spectrum is such that ${\widetilde m} \gg
m_{1/2}$, like in the case of split SUSY, there are no Q-balls as $K >
0$.

The potential can also be much flatter $\propto \textrm{ln} |\phi|$ at
large field values. This happens in models of gauge-mediated SUSY
breaking~\cite{gmsb}, and can arise from thermal
corrections~\cite{and}.  The important point in any case is that the
scalar field profile within a Q-ball is such that the field value is
maximum at the center $\varphi_0$ and decreases towards the
surface. This implies that for $g \varphi_0 \geq {\widetilde m}$, with
$g$ being a typical coupling of the $\phi$ field to other fields, the
Q-ball decays through its surface as decay inside the Q-ball is not
energetically allowed~\cite{cohenetal}. Note that for a typical gauge
or Yukawa coupling $g \varphi_0 \geq {\widetilde m}$ if $\varphi_0 \gg
{\widetilde m}$. The decay rate of Q-ball which contains a
(baryonic/leptonic) charge $Q$ is in this case given
by~\cite{cohenetal}
\beq \label{qballrate}
{d Q \over d t} \leq {\omega^3 A \over 192 \pi^2},
\eeq
where $\omega \simeq {\widetilde m}$ and $A = 4 \pi R^2_{\rm Q}$ is the
surface area of the Q-ball. For example, consider Q-ball formation for
the potential given in~(\ref{gravqball}). A Q-ball with total charge
$Q$ then has a decay lifetime~\cite{enqvistetal}
\beq \label{qballife}
\tau_{\rm Q} \gsim \left({|K| \over 0.03}\right) \left({1~{\rm TeV} \over 
{\widetilde m}}\right) \left({Q \over 10^{20}}\right) \times 
10^{-7}~{\rm sec},
\eeq
which corresponds to the decay temperature
\beq \label{qballtemp}
T_{\rm d} \lsim \left({0.03 \over |K|}\right)^{1/2} \left({{\widetilde m} 
\over 1~{\rm TeV}}\right)^{1/2} \left({10^{20} \over Q}\right)^{1/2} 
\times 2~{\rm GeV}.
\eeq
Here we have used $\tau^{-1}_{\rm d} \sim \left(T^2_{\rm d}/M_{\rm P}
\right)$. The total baryonic/leptonic charge $Q$ of a Q-ball is given
by the multiplication of baryon/lepton number carried by the flat
direction and the total number of zero-mode quanta inside the Q-ball.
This, after using~(\ref{decrate}), leads to
\begin{eqnarray} \label{qballdecay}
{\rm Helicity} \pm {1 \over 2}: \left({n_{3/2} \over s}\right)_{\rm
Q-ball} &\sim & \left({{\widetilde m} \over m_{3/2}}\right)^2
\left({{\widetilde m} \over 1~{\rm TeV}}\right)^2
\left({Q \over 10^{20}}\right) 
\left({n_{\rm B} \over s}\right) 2.5 \times 10^{-13} \,, \nonumber \\ 
{\rm Helicity} \pm {3 \over 2}: \left({n_{3/2} \over s}\right)_{\rm
Q-ball} &\sim & \left({{\widetilde m} \over 1~{\rm TeV}}\right)^2
\left({Q \over 10^{20}}\right) 
\left({n_{\rm B} \over s}\right) 2.5 \times 10^{-13} \,,
\end{eqnarray}
Obviously $\left(n_{3/2}/s \right)_{\rm Q-ball} \leq \left(n_{\rm B}/s
\right)$, since the decay of each quanta inside the Q-ball can at most 
produce one gravitino. This implies that gravitinos produced from the decay of
Q-balls will not dominate the Universe if the decay generates a baryon
asymmetry $\left(n_{\rm B}/s \right) \simeq 10^{-10}$, see
Eq.~(\ref{dom}). However, the situation will be different for larger
Q-balls which yield an asymmetry $\gg 10^{-10}$. If gravitinos from
Q-ball decay dominate the Universe, they will dilute the baryon
asymmetry.  The final asymmetry will then have the correct size, see
Eqs.~(\ref{dilution}) and~(\ref{qballdecay}), if
\beq \label{qballbaryon}
Q \sim \left({m_{3/2} \over 10^5~{\rm GeV}}\right)^{1/2} 
\left({1~{\rm TeV} \over {\widetilde m}}\right)^2 \left({m_{3/2} 
\over {\widetilde m}}\right)^2 \times 10^{35}.
\eeq
Assuming that the Q-balls do not dominate the energy density of the
Universe, we must have $\left(n_{\rm B}/s \right) < \left(3 T_{\rm
d}/4 {\widetilde m}\right)$. The necessary condition for gravitino
domination, after using Eqs.~(\ref{dom}),~(\ref{qballtemp})
and Eq.~(\ref{qballdecay}), is then obtained to be
\beq \label{qballdom}
Q > \left({m_{3/2} \over 10^5~{\rm GeV}}\right)
\left({m_{3/2} \over {\widetilde m}}\right)^4
\left({1~{\rm TeV} \over {\widetilde m}}\right)^3 2.5 \times 10^{35}.
\eeq
For ${\widetilde m} \gg 1$ TeV and $m_{3/2} \ll {\widetilde m}$, this
lower bound on the Q-ball charge is compatible with the value in
Eq.~(\ref{qballbaryon}) required for successful baryogenesis. Note
that the Q-ball decay temperature $T_{\rm d}$ may be smaller than the
LSP freeze-out temperature $T_{\rm f}$. In this case Q-ball decay can
be dangerous as three LSP per baryon number will be produced. However,
this will not lead to problem so long as the gravitino decay
temperature $T_{3/2} < T_{\rm d}$ (which is typically the case) and
the condition for efficient LSP annihilation in Eq.~(\ref{eff}) is
satisfied.  The Q-balls will dominate the Universe if they carry a
very large charge. The initial baryon asymmetry released by the
Q-balls then has a simple expression $\left(n_{\rm B}/s \right) \sim
\left(T_{\rm d}/{\widetilde m}\right)$.

The results in Eqs.~(\ref{qballife}),~(\ref{qballtemp}),
~(\ref{qballdecay}),~(\ref{qballbaryon}) and~(\ref{qballdom}) are
valid for the potential given in Eq.~(\ref{gravqball}). The same steps
(though more involved) can be followed to obtain similar results for
logarithmic potentials as in the case of gauge-mediated models. The
remarkable point in all cases is that the Q-ball decay lifetime
increases with its charge, implying a more efficient production of
gravitinos from the Q-ball decay.  This leads to an attractive
solution that the large baryon asymmetry released by the Q-ball decay
can be naturally diluted by gravitinos produced in the same process.

Finally, we shall notice that in models of running mass inflation even
the inflaton condensate could fragment into non-topological
solitons~\cite{Kasuya}. In that case Q-balls would naturally dominate
the Universe. We do not discuss such possibility here.

\section{Summary and Conclusions}

In this paper we have investigated cosmological consequences of models
with superheavy gravitinos and/or sfermions. A priori there is no
fundamental reason which fixes the scale of SUSY breaking. Models with
weak-scale SUSY breaking in the observable sector have the promise to
solve the hierarchy problem. However this may not necessarily be the
case and the SUSY breaking scale can turn out to be very high. Under
general circumstances, arbitrarily heavy gravitino mass $m_{3/2}$
and/or sfermion masses ${\widetilde m}$ are quite
plausible. Therefore, inspired from the recent models of large scale
SUSY breaking, it becomes pertinent to re-examine the cosmological 
and phenomenological consequences.

Gravitino are produced through various processes in the early
Universe.  Scatterings of gauge and gaugino quanta in thermal bath,
sfermion decays and the inflaton decay are the main sources for
gravitino production.  The main results are presented in
Eqs.~(\ref{scattering}),~(\ref{decay}) and~(\ref{phidecay}). Sfermion
decays usually dominate when the reheating temperature $T_{\rm R} \geq
{\widetilde m} > m_{3/2}$. On the other hand, the contribution from
the inflaton decay dominates when $T_{\rm R} \ll {\widetilde m}$.

Gravitinos which are heavier than $50$ TeV decay before primordial
nucleosynthesis, and hence are not subject to BBN bounds. However,
each gravitino produces one LSP upon its decay. Hence, in models with
conserved $R$-parity, the abundance of {\it supermassive} gravitinos
is constrained by the dark matter limit. Indeed, efficient
annihilation of LSPs produced in gravitino decay sets a lower bound on
$m_{3/2}$. When this lower bound is saturated, gravitino decay can
successfully produce non-thermal LSP dark matter.

This is also valid in a gravitino-dominated Universe, which happens
when gravitinos are produced very abundantly. However, for $m_{3/2}
\geq 50$ TeV, gravitino domination cannot rescue a scenario where
thermal LSP abundance at the freeze-out exceeds the dark matter
bound. For a Wino- or Higgsino-like LSP this is the case when the LSP
mass $m_{\chi} > 2$ TeV. The reason is that in this case gravitino
decay, while diluting the thermal abundance, leads to non-thermal
overproduction of LSPs. Therefore, if $R$-parity is conserved,
gravitinos should never dominate in models with such heavy LSPs.  The
results for gravitino production in conjunction with the constraints
from the dark matter bound and LSP annihilation are summarized in
Eqs.~(\ref{after}),~(\ref{eff}),~(\ref{nosfermdom})
and~(\ref{sfermdom}). Figs.~(1) and~(2) depict the acceptable parts of
the ${\widetilde m}-m_{3/2}$ plane for successful scenarios of
gravitino non-domination and domination respectively.

We discussed some specific scenarios of baryogenesis in the presence
of supermassive gravitinos. The parameter space for thermal
leptogenesis is substantially relaxed when $m_{3/2} \geq 50$ TeV, as a
considerably larger reheating temperature $T_{\rm R}$ and/or right-handed
(s)neutrino mass $M_1$ will be allowed. This, however, implies that
gravitinos can also be efficiently produced for a wide range of
sfermion masses. Since gravitino decay takes place after the
completion of leptogenesis, unless they are extremely heavy, the
generated baryon asymmetry will be diluted in a gravitino-dominated
Universe. Successful leptogenesis then requires (much) larger
right-handed (s)neutrino masses than usual. However, it is known that
thermal leptogenesis fails for $M_1 \geq 10^{14}$ GeV since lepton
number violating scatterings in this case erase the generated
asymmetry. This leads to additional constraints on the ${\widetilde
m}-m_{3/2}$ parameter space in case of gravitino domination. Our
results are summarized in Eqs.~(\ref{domasym}),~(\ref{suclep}) and
Fig.~(4).

We also considered late time baryogenesis from supersymmetric flat
directions via the Affleck-Dine mechanism. Thermal effects which can
trigger early oscillations of the flat direction condensate, thus
suppressing the generate asymmetry, tend to be less important for
${\widetilde m} \gg 1$ TeV.  A large expectation value for the
condensate at the onset of its oscillations usually leads to a baryon
asymmetry $\left(n_{\rm B}/s \right) \gg 10^{-10}$, as well as a large
reheating temperature $T_{\rm R} \geq {\widetilde m}$. Gravitinos
produced from sfermion decays can then dominate the Universe and
dilute the initially large asymmetry down to acceptable values. The
main results in Eqs.(\ref{barentr2}), ~(\ref{barentr3}) are depicted
in Figs~(5),~(6).

There is even a closer connection between large baryon asymmetry and 
efficient gravitino production when oscillations of the flat direction 
condensate fragment into Q-balls (as happen in many cases). Q-balls decay 
(much) later than the homogeneous condensate, and the larger the 
baryonic/leptonic charge they carry the longer their decay lifetime. 
Hence the decay of large Q-balls is a natural source for 
copious production of gravitinos which can even dominate the Universe and 
dilute the large baryon asymmetry released by Q-balls. We explicitly 
demonstrated this for a potential with logarithmic corrections, 
Eqs.~(\ref{qballbaryon}) and~(\ref{qballdom}), but the same conclusions hold 
for other types of flat potentials.

To conclude, models with superheavy gravitinos and/or sfermions have
very interesting cosmological consequences. These models can naturally
give rise to a large gravitino abundance in the early Universe. This,
contrary to models with a weak scale gravitino mass, can turn to a
virtue and lead to successful production of dark matter and baryon
asymmetry generation.


\section{Acknowledgments}

The work of R.A. is supported by the National Sciences and Engineering
Research Council. S.P thanks the kind hospitality of NORDITA and NBI during
part of the course of the present work.



\begin{thebibliography}{99}

\bibitem{infl} 
For reviews on inflation, see: A. D. Linde, {\it Particle Physics and
Inflationary Cosmology}, Harwood, Chur, Switzerland (1990).

\bibitem{WMAP} C. L. Bennett, et.al. First Year Wilkinson Microwave
Anisotropy Probe (WMAP) Observations: Preliminary Maps and Basic
Results, Astrophys. J. Suppl. {\bf 148}, 1 (2003)[arXiv:astro-ph/0302207]

\bibitem{Dolgov}
A.~Albrecht, P.~J.~Steinhardt, M.~S.~Turner and F.~Wilczek,
Phys.\ Rev.\ Lett.\  {\bf 48}, 1437 (1982).
A.~D.~Dolgov and A.~D.~Linde,
Phys.\ Lett.\ B {\bf 116}, 329 (1982).
L.~F.~Abbott, E.~Farhi and M.~B.~Wise,
Phys.\ Lett.\ B {\bf 117}, 29 (1982).

\bibitem{Brandenberger}
J.~H.~Traschen and R.~H.~Brandenberger,
Phys.\ Rev.\ D {\bf 42}, 2491 (1990).
Y.~Shtanov, J.~H.~Traschen and R.~H.~Brandenberger,
Phys.\ Rev.\ D {\bf 51}, 5438 (1995)[arXiv:hep-ph/9407247].

\bibitem{kls}
L.~Kofman, A.~D.~Linde and A.~A.~Starobinsky,
Phys. Rev. Lett. {\bf 73}, 3195 (1994)[arXiv:hep-th/9405187].
L.~Kofman, A.~D.~Linde and A.~A.~Starobinsky,
Phys.\ Rev.\ D {\bf 56}, 3258 (1997)[arXiv:hep-ph/9704452].

\bibitem{chm}
D.~Cormier, K.~Heitmann and A.~Mazumdar, Phys.\ Rev.\ D {\bf 65}, 083521 (2002)
[arXiv:hep-ph/0105236].

\bibitem{Greene}
P.~B.~Greene and L.~Kofman,
Phys.\ Lett.\ B {\bf 448}, 6 (1999)[arXiv:hep-ph/9807339].
J.~Baacke, K.~Heitmann and C.~Patzold,
Phys.\ Rev.\ D {\bf 58}, 125013 (1998)[arXiv:hep-ph/9806205].

\bibitem{bbn}
For a review on BBN, see: K. A. Olive, G. Steigman and T. P. Walker, 
Phys. Rept. {\bf 333}, 389 (2000)[arXiv:astro-ph/9905320]. 

\bibitem{subir}
For a review, see: S. Sarkar, 
Rept. Prog. Phys. {\bf 59}, 1493 (1996)[arXiv:hep-ph/9602260].

\bibitem{lowt}
M.~Kawasaki, K.~Kohri and N.~Sugiyama,
Phys.\ Rev.\ D {\bf 62}, 023506 (2000)[arXiv:astro-ph/0002127].
S.~Hannestad,
Phys.\ Rev.\ D {\bf 70}, 043506 (2004)[arXiv:astro-ph/0403291].

\bibitem{Nilles}
For a review on SUSY, see: H.~P.~Nilles,
Phys.\ Rept.\  {\bf 110}, 1 (1984).

\bibitem{gkm}
T. Gherghetta, C. F. Kolda and S. P. Martin, Nucl. Phys. B {\bf 468}, 35 
(1996)[arXiv:hep-ph/9510370].   

\bibitem{Kari}
K.~Enqvist and A.~Mazumdar,
Phys.\ Rept.\  {\bf 380}, 99 (2003)[arXiv:hep-ph/0209244].

\bibitem{Jungman}
For a review on supersymmetric dark matter, see: G.~Jungman, M.~Kamionkowski 
and K.~Griest,
Phys.\ Rept.\  {\bf 267}, 195 (1996)
[arXiv:hep-ph/9506380]. 
P. Gondolo, J. Edsjo, P. Ullio, L. Bergstrom, M. Schellke and E. A. Baltz, 
JCAP {\bf 0407}, 008 (2004)
[arXiv:astro-ph/0406204].

\bibitem{gr}
G.~F.~Giudice and R.~Rattazzi,
Phys.\ Rept.\  {\bf 322}, 419 (1999)
[arXiv:hep-ph/9801271].

\bibitem{anomaly}
G. F. Giudice, M. A. Luty, H. Murayama and R. Rattazzi, 
JHEP {\bf 9812}, 027 (1998)
[arXiv:hep-ph/9810442].

\bibitem{Landscape}
R.~Bousso and J.~Polchinski,
JHEP {\bf 0006}, 006 (2000)
[arXiv:hep-th/0004134].
S.~Kachru, R.~Kallosh, A.~D.~Linde and S.~P.~Trivedi,
Phys.\ Rev.\ D {\bf 68}, 046005 (2003)
[arXiv:hep-th/0301240].
S.~Kachru, R.~Kallosh, A.~D.~Linde, J.~Maldacena, L.~McAllister and 
S.~P.~Trivedi,
JCAP {\bf 0310}, 013 (2003)
[arXiv:hep-th/0308055].
L.~Susskind,
arXiv:hep-th/0302219.
M.~R.~Douglas,
JHEP {\bf 0305}, 046 (2003)
[arXiv:hep-th/0303194].
S.~Ashok and M.~R.~Douglas,
JHEP {\bf 0401}, 060 (2004)
[arXiv:hep-th/0307049].
S.~B.~Giddings, S.~Kachru and J.~Polchinski,
Phys.\ Rev.\ D {\bf 66}, 106006 (2002)
[arXiv:hep-th/0105097].
J.~P.~Conlon and F.~Quevedo,
JHEP {\bf 0410}, 039 (2004)
[arXiv:hep-th/0409215].
T.~Banks, M.~Dine and E.~Gorbatov,
JHEP {\bf 0408}, 058 (2004)
[arXiv:hep-th/0309170].

\bibitem{Douglas}
M.~R.~Douglas,
arXiv:hep-ph/0401004.
L.~Susskind,
arXiv:hep-th/0405189.
M.~R.~Douglas,
Comptes Rendus Physique {\bf 5}, 965 (2004)
[arXiv:hep-th/0409207].
E.~Silverstein,
arXiv:hep-th/0407202.
M.~Dine, E.~Gorbatov and S.~Thomas,
arXiv:hep-th/0407043.

\bibitem{Cliff0}
C.~P.~Burgess, R.~Easther, A.~Mazumdar, D.~F.~Mota and T.~Multamaki,
arXiv:hep-th/0501125.

\bibitem{split}
N.~Arkani-Hamed and S.~Dimopoulos,
arXiv:hep-th/0405159.
G.~F.~Giudice and A.~Romanino,
Nucl.\ Phys.\ B {\bf 699}, 65 (2004)
[Erratum-ibid.\ B {\bf 706}, 65 (2005)]
[arXiv:hep-ph/0406088].

\bibitem{adgr}
N.~Arkani-Hamed, S.~Dimopoulos, G.~F.~Giudice and A.~Romanino,
arXiv:hep-ph/0409232.

\bibitem{models}
K. S. Babu, T. Enkhbat and B. Mukhopadhyaya, arXiv:hep-ph/0501079. 
M. Drees, arXiv:hep-ph/0501106. 
N. Haba and N. Okada, arXiv:hep-ph/0502213.  

\bibitem{inverted}
A. G. Cohen, D. B. Kaplan and A. E. Nelson, Phys. Lett. B {\bf 388}, 588 (1996)
[arXiv:hep-ph/9607394]. 
J. A. Bagger, J. L. Feng, N. Polonsky and R. J. Zhang, Phys. Lett. B 
{\bf 473}, 264 (2000)
[arXiv:hep-ph/9911255]. 


\bibitem{thermal}
M. Y. Khlopov and A. D. Linde, 
Phys. Lett. B {\bf 138}, 265 (1984).
J. R. Ellis, J. E. Kim and D. V. Nanopoulos,
Phys. Lett. B {\bf 145}, 181 (1984).
J.~R.~Ellis, D.~V.~Nanopoulos, K.~A.~Olive and S.~J.~Rey,
Astropart.\ Phys.\  {\bf 4}, 371 (1996)
[arXiv:hep-ph/9505438].
M.~Bolz, A.~Brandenburg and W.~Buchm\"uller,
Nucl.\ Phys.\ B {\bf 606}, 518 (2001)
[arXiv:hep-ph/0012052].

\bibitem{gmsb}
T.~Moroi, H.~Murayama and M.~Yamaguchi,
Phys.\ Lett.\ B {\bf 303}, 289 (1993).
A.~de Gouvea, T.~Moroi and H.~Murayama,
Phys.\ Rev.\ D {\bf 56}, 1281 (1997)
[arXiv:hep-ph/9701244].

\bibitem{nop}
H. P. Nilles, K. A. Olive and M. Peloso, 
Phys. Lett. B {\bf 522}, 304 (2001)
[arXiv:hep-ph/0107212].

\bibitem{ajm}
R. Allahverdi, A. Jokinen and A. Mazumdar, 
Phys. Rev. D {\bf 71}, 043505 (2005)
[arXiv:hep-ph/0410169].

\bibitem{aem}
R.~Allahverdi, K.~Enqvist and A.~Mazumdar,
Phys.\ Rev.\ D {\bf 65}, 103519 (2002)
[arXiv:hep-ph/0111299].

\bibitem{non-pert1}
A.~L.~Maroto and A.~Mazumdar,
Phys.\ Rev.\ Lett.\  {\bf 84}, 1655 (2000)
[arXiv:hep-ph/9904206].

\bibitem{non-pert2}
R.~Kallosh, L.~Kofman, A.~D.~Linde and A.~Van Proeyen,
Phys.\ Rev.\ D {\bf 61}, 103503 (2000)
[arXiv:hep-th/9907124].
R.~Kallosh, L.~Kofman, A.~D.~Linde and A.~Van Proeyen,
Class.\ Quant.\ Grav.\  {\bf 17}, 4269 (2000)
[arXiv:hep-th/0006179].
A. L. Maroto and J. R. Pelaez, Phys. Rev. D {\bf 62}, 023518 (2000);
G.~F.~Giudice, A.~Riotto and I.~I.~Tkachev,
JHEP {\bf 9911}, 036 (1999)
[arXiv:hep-ph/9911302].
G.~F.~Giudice, I.~I.~Tkachev and A.~Riotto,
JHEP {\bf 9908}, 009 (1999)
[arXiv:hep-ph/9907510].
M.~Bastero-Gil and A.~Mazumdar,
Phys.\ Rev.\ D {\bf 62}, 083510 (2000)
[arXiv:hep-ph/0002004].

\bibitem{non-pert3}
R.~Allahverdi, M.~Bastero-Gil and A.~Mazumdar,
Phys.\ Rev.\ D {\bf 64}, 023516 (2001)
[arXiv:hep-ph/0012057].

\bibitem{nps}
H.~P.~Nilles, M.~Peloso and L.~Sorbo,
Phys.\ Rev.\ Lett.\  {\bf 87}, 051302 (2001)
[arXiv:hep-ph/0102264].
H.~P.~Nilles, M.~Peloso and L.~Sorbo,
JHEP {\bf 0104}, 004 (2001)
[arXiv:hep-th/0103202].

\bibitem{no-scale}
J. R. Ellis, C. Kounnas and D. V. Nanopoulos, Phys. Lett. B {\bf 143}, 410 
(1984).

\bibitem{cefo}
R.~H.~Cyburt, J.~R.~Ellis, B.~D.~Fields and K.~A.~Olive,
Phys.\ Rev.\ D {\bf 67}, 103521 (2003)
[arXiv:astro-ph/0211258].

\bibitem{kkm} 
M.~Kawasaki, K.~Kohri and T.~Moroi,
arXiv:astro-ph/0402490.
M.~Kawasaki, K.~Kohri and T.~Moroi,
arXiv:astro-ph/0408426.

\bibitem{plumacher}
M. Pl\"umacher, Z. Phys. C {\bf 74}, 549 (1997)
[arXiv:hep-ph/9604229]. 
M. Plumacher, Nucl. Phys. B {\bf 530}, 207 (1998)
[arXiv:hep-ph/9704231]. 
W. Buchm\"uller and M. Pl\"umacher, Phys. Rept. {\bf 320}, 329 (1999)
[arXiv:hep-ph/9904310]. 
W. Buchm\"uller and M. Pl\"umacher, Int. J. Mod. Phys. A {\bf 15}, 5047 (2000)
[arXiv:hep-ph/0007176].

\bibitem{ad}
I. Affleck and M. Dine, Nucl. Phys. B {\bf 249}, 361 (1985).

\bibitem{cohen}
A.~G.~Cohen, S.~R.~Coleman, H.~Georgi and A.~Manohar,
Nucl.\ Phys.\ B {\bf 272}, 301 (1986).

\bibitem{moroi}
T. Moroi, 
arXiv:hep-ph/9503210.

\bibitem{kyy}
K.~Kohri, M.~Yamaguchi and J.~Yokoyama,
Phys.\ Rev.\ D {\bf 70}, 043522 (2004)
[arXiv:hep-ph/0403043]. 
K, Kohri, M. Yamaguchi and J. Yokoyama, arXiv:hep-ph/0502211.

\bibitem{ad4}
R.~Allahverdi and M.~Drees,
Phys. Rev. D {\bf 70}, 123522 (2004)
[arXiv:hep-ph/0408289].

\bibitem{thermalization}
S. Davidson and S. Sarkar, 
JHEP {\bf 0011}, 012 (2000)
[arXiv:hepph/0009078]. 
R. Allahverdi and M. Drees,  
Phys. Rev. D {\bf 66}, 063513 (2002)
[arXiv:hep-ph/0205246]. 
P. Jaikumar and A. Mazumdar, 
Nucl. Phys. B {\bf 683}, 264 (2004)
[arXiv:hep-ph/0212265].


\bibitem{ad1}
R. Allahverdi and M. Drees, 
Phys. Rev. Lett. {\bf 89}, 091302 (2002)
[arXiv:hep-ph/0203118].


\bibitem{abgp}
R. Allahverdi, C. Bird, S. Groot Nibbelink and M. Pospelov, 
Phys. Rev. D {\bf 69}, 045004 (2004)
[arXiv:hep-ph/0305010].

\bibitem{kamionkowski}
X. Chen, M. Kamionkowski and X. Zhang, 
Phys. Rev. D {\bf 64}, 021302 (2001)
[arXiv:astro-ph/0103452].

\bibitem{ggw}
T. Gherghetta, G. F. Giudice and J. D. Wells, Nucl. Phys. B {\bf 559}, 27 
(1999)
[arXiv:hep-h/9904378].

\bibitem{cfo}
R.~H.~Cyburt, B.~D.~Fields and K.~A.~Olive,
Phys.\ Lett.\ B {\bf 567}, 227 (2003)
[arXiv:astro-ph/0302431].

\bibitem{sakharov}
A. D. Sakharov, 
JETP Lett. B {\bf 91}, 24 (1967). 

\bibitem{krs}
V. Kuzmin, V. A. Rubakov, and M. E. Shaposhnikov, 
Phys. Lett. B {\bf 155}, 36 (1985).

\bibitem{khlebnikov}
S. Yu. Khlebnikov and M. E. Shaposhnikov, Nucl. Phys. B {\bf 308}, 885
(1985). 

\bibitem{seesaw}
P.~Minkowski,
Phys.\ Lett.\ B {\bf 67}, 421 (1977);
M. Gell-Mann, P. Ramond and R. Slansky, in {\it Supergravity},
eds. P. van Nieuwenhuizen and D. Z. Freedman (North Holland, 1979);
T. Yanagida, Proceedings of {\it Workshop on
Unified Theory and Baryon number in the Universe}, eds.
O. Sawada and A. Sugamoto (KEK, Tsukuba, 1979);
R. N. Mohapatra and G. Senjanovic, Phys. Rev. Lett. {\bf 44}, 912
(1980).

\bibitem{fy}
M. Fukugita and T. Yanagida, Phys. Lett. B {\bf 174}, 45 (1986).

\bibitem{luty}
M. A. Luty, Phys.\ Rev.\ D {\bf 45}, 455 (1992).

\bibitem{one-loop}
M. Flanz, E. A. Paschos and U. Sarkar, Phys. Lett. B {\bf 345} 
248 (1995)
[arXiv:hep-ph/9411366]. 
L. Covi, E. Roulet and F. Vissani, Phys. Lett. B {\bf 384}, 
169 (1996)
[arXiv:hep-ph/9605319]. 
W. Buchm\"uller and M. Pl\"umacher, Phys. Lett. B {\bf 431}, 
354 (1998)
[arXiv:hep-ph/9710460].  


\bibitem{buchmuller}
W. Buchm\"uller, P. Di Bari and M. Pl\"umacher, Phys. Lett. B {\bf 547}, 128 
(2002)
[arXiv:hep-ph/0209301]. 
W. Buchm\"uller, P. Di Bari and M. Pl\"umacher, Nucl. Phys. B {\bf 665}, 
445 (2003)
[arXiv:hep-ph/0302092]. 
W. Buchm\"uller, P. Di Bari and M. Pl\"umacher, Annals. Phys. {\bf 315}, 
305 (2005)
[arXiv:hep-ph/0401240]. 
W. Buchm\"uller, P. Di Bari and M. Pl\"umacher, New. J. Phys. {\bf 6}, 105 
(2004)
[arXiv:hep-ph/0406014].

\bibitem{sacha}
S. Davidson, JHEP {\bf 0303}, 037 (2003)
[arXiv:hep-ph/0302075].

\bibitem{gnrrs}
G. F. Giudice, A. Notari, M. Raidal, A. Riotto and A. Strumia,
Nucl. Phys. B {\bf 685}, 89 (2004)
[arXiv:hep-ph/0310123].  

\bibitem{pilaftsis}
A. Pilaftsis, Phys. Rev. D {\bf 56}, 5431 (1997)
[arXiv:hep-ph/9707235]. 
A. Pilaftsis, Int. J. Mod. Phys. A {\bf 14}, 1811 (1999)
[arXiv:hep-ph/9812256]. 
A. Pilaftsis, arXiv:hep-ph/0408103. 
A. Pilaftsis and T. E. J. Underwood, Nucl. Phys. B {\bf 692}, 303 (2004)
[arXiv:hep-ph/0309342]. 
T. Hambye, J. March-Russell and S. M. West, JHEP {\bf 0407}, 070 (2004)
[arXiv:hep-ph/0403183].   

\bibitem{strumia}
T. Hambye, Y. Lin, A. Notari, M. Papucci and A. Strumia, Nucl. Phys. B 
{\bf 695}, 169 (2004)
[arXiv:hep-ph/0312203]. 
M. Raidal, A. Strumia and K. Turzynski, arXiv:hep-ph/0408015.

\bibitem{reheat}
G. Lazarides and Q. Shafi, Phys. Lett. B {\bf 258}, 305 (1991). 
K. Kumekawa, T. Moroi and T. Yanagida, Prog. Theor. Phys. {\bf 92},
437 (1994)
[arXiv:hep-ph/9405337]. 
T. Asaka, K. Hamaguchi, M. Kawasaki and T. Yanagida, Phys. Lett. B {\bf 464}, 
12 (1999)
[arXiv:hep-ph/9906366]. 
T. Asaka, K. Hamaguchi, M. Kawasaki and T. Yanagida, Phys. Rev. D {\bf 61}, 
083512 (2000)
[arXiv:hep-ph/9907559].

\bibitem{gprt}
G. F. Giudice, M. Peloso, A. Riotto and I. I. Tkachev, JHEP {\bf 9908}, 014 
(1999)
[arXiv:hep-ph/9905242]. 


\bibitem{off-shell}
L. Bento and Z. Berezhiani, Phys. Rev. Lett. {\bf 87}, 231304 (2001)
[arXiv:hep-ph/0107281]. 
R. Allahverdi and A. Mazumdar, Phys. Rev. D {\bf 67}, 023509 (2003)
[arXiv:hep-ph/0208268]. 
T. Dent, G. Lazarides and R. Ruiz de Austri, Phys. Rev. D {\bf 69}, 
075012 (2004)
[arXiv:hep-ph/0312033]. 
T. Dent, G. Lazarides and R. Ruiz de Austri, arXiv:hep-ph/0503235.

\bibitem{cdo}
B. A. Campbell, S. Davidson and K. A. Olive, 
Nucl. Phys. B {\bf 399}, 111 (1993)
[arXiv:hep-ph/9302223]. 

\bibitem{bdps}
L. Boubekeur, S. Davidson, M. Peloso and L. Sorbo, Phys. Rev. D {\bf 67}, 
043515 (2003)
[arXiv:hep-ph/0209256].

\bibitem{my}
H. Murayama and T. Yanagida, Phys. Lett. B {\bf 322}, 349 (1994)
[arXiv:hep-ph/9310297].

\bibitem{hmy}
K. Hamaguchi, H. Murayama and T. Yanagida, Phys. Rev. D {\bf 65},
043512 (2002)
[arXiv:hep-ph/0109030].  

\bibitem{bmp}
Z. Berezhiani, A. Mazumdar and A. P\'erez-Lorenzana, 
Phys. Lett. B {\bf 518}, 282 (2001)
[arXiv:hep-ph/0107239]. 

\bibitem{adm}
R. Allahverdi, B. Dutta and A. Mazumdar, 
Phys. Rev. D {\bf 67}, 123515 (2003)
[arXiv:hep-ph/0301184].

\bibitem{ad3}
R.~Allahverdi and M.~Drees,
Phys.\ Rev.\ D {\bf 69}, 103522 (2004)
[arXiv:hep-ph/0401054].

\bibitem{soft}
Y. Grossman, T. Kashti, Y. Nir and E. Roulet, Phys. Lett. Rev. {\bf 91}, 
251801 (2003)
[arXiv:hep-ph/0307081]. 
Y. Grossman, T. Kashti, Y. Nir and E. Roulet, JHEP {\bf 0411}, 080 (2004)
[arXiv:hep-ph/0407063]. 
G. D'Ambrosio, G. F. Giudice and M. Raidal, Phys. Lett. B {\bf 575}, 
75 (2003)
[arXiv:hep-ph/0308031]. 
L. Boubekeur, T. Hambye and G. Senjanovic, Phys. Rev. Lett. {\bf 93}, 
111601 (2004)
[arXiv:hep-ph/0404038].   


\bibitem{di}
S. Davidson and A. Ibarra, Phys. Lett. B {\bf 535}, 25 (2002)
[arXiv:hep-ph/0202239]. 


\bibitem{CPVconn1}
S.~Davidson and A.~Ibarra,
JHEP {\bf 0109}, 013 (2001)
[arXiv:hep-ph/0104076].
G.~C.~Branco {\it et al.}, 
Nucl.\ Phys.\ B {\bf 617}, 475 (2001)
[arXiv:hep-ph/0107164].
J.~R.~Ellis {\it et al.}, 
Nucl.\ Phys.\ B {\bf 621}, 208 (2002)
[arXiv:hep-ph/0109125].
G.~C.~Branco {\it et al.}, 
Nucl.\ Phys.\ B {\bf 640}, 202 (2002)
[arXiv:hep-ph/0202030].
J.~R.~Ellis and M.~Raidal,
Nucl.\ Phys.\ B {\bf 643}, 229 (2002)
[arXiv:hep-ph/0206174].
P.~H.~Frampton, S.~L.~Glashow and T.~Yanagida,
Phys.\ Lett.\ B {\bf 548}, 119 (2002)
[arXiv:hep-ph/0208157].
S.~Davidson and A.~Ibarra,
Nucl.\ Phys.\ B {\bf 648}, 345 (2003)
[arXiv:hep-ph/0206304].
S.~Kaneko and M.~Tanimoto,
Phys.\ Lett.\ B {\bf 551}, 127 (2003)
[arXiv:hep-ph/0210155].
G.~C.~Branco {\it et al.}, 
Phys.\ Rev.\ D {\bf 67}, 073025 (2003)
[arXiv:hep-ph/0211001].
S.~Pascoli, S.~T.~Petcov and W.~Rodejohann,
Phys.\ Rev.\ D {\bf 68}, 093007 (2003)
[arXiv:hep-ph/0302054].
S.~Kaneko, M.~Katsumata and M.~Tanimoto,
JHEP {\bf 0307}, 025 (2003)
[arXiv:hep-ph/0305014].
S.~Davidson and R.~Kitano,
JHEP {\bf 0403}, 020 (2004)
[arXiv:hep-ph/0312007].
A.~Ibarra and G.~G.~Ross,
Phys.\ Lett.\ B {\bf 591}, 285 (2004)
[arXiv:hep-ph/0312138].
S. Pascoli, Mod.\ Phys.\ Lett.\ A {\bf 20}, 477 (2005).
M.~Bando {\it et al.}, 
arXiv:hep-ph/0405071.


\bibitem{fhy}
M. Fujii, K. Hamaguchi and T. Yanagida, Phys. Rev. D {\bf 65}, 115012 (2002)
[arXiv:hep-ph/0202210].

\bibitem{denspert}
K.~Enqvist, S.~Kasuya and A.~Mazumdar,
Phys.\ Rev.\ Lett.\  {\bf 90}, 091302 (2003)
[arXiv:hep-ph/0211147].
K.~Enqvist, A.~Jokinen, S.~Kasuya and A.~Mazumdar,
Phys.\ Rev.\ D {\bf 68}, 103507 (2003)
[arXiv:hep-ph/0303165].
G. Dvali, A. Gruzinov and M. Zaldarriaga, Phys. Rev. D {\bf 69}, 023505 (2004)
[arXiv:astro-ph/0303591]; 
K.~Enqvist, A.~Mazumdar and M.~Postma,
Phys.\ Rev.\ D {\bf 67}, 121303 (2003)
[arXiv:astro-ph/0304187].
A.Mazumdar and M. Postma,
Phys.\ Lett.\ B {\bf 573}, 5 (2003)
[Erratum-ibid.\ B {\bf 585}, 295 (2004)]
[arXiv:astro-ph/0306509]. 
R. Allahverdi, Phys. Rev. D {\bf 70}, 043507 (2004)
[arXiv:astro-ph/0403351].  
K.~Enqvist, S.~Kasuya and A.~Mazumdar,
Phys.\ Rev.\ Lett.\  {\bf 93}, 061301 (2004)
[arXiv:hep-ph/0311224].
K.~Enqvist, A.~Mazumdar and A.~Perez-Lorenzana,
arXiv:hep-th/0403044.

\bibitem{mag}
K.~Enqvist, A.~Jokinen and A.~Mazumdar,
JCAP {\bf 0411}, 001 (2004)
[arXiv:hep-ph/0404269].

\bibitem{averdi}
R.~Allahverdi, R.~Brandenberger and A.~Mazumdar,
Phys. Rev. D {\bf 70}, 083535 (2004)[
arXiv:hep-ph/0407230].

\bibitem{drt}
M.~Dine, L.~Randall and S.~Thomas,
Nucl.\ Phys.\ B {\bf 458}, 291 (1996)
[arXiv:hep-ph/9507453].

\bibitem{Gaillard}
M.~K.~Gaillard, H.~Murayama and K.~A.~Olive,
Phys.\ Lett.\ B {\bf 355}, 71 (1995)
[arXiv:hep-ph/9504307].

\bibitem{manuel}
R. Allahverdi, M. Drees and A. Mazumdar, Phys. Rev. D {\bf 65}, 065010 (2002)
[arXiv:hep-ph/0108225].  

\bibitem{ace} 
R. Allahverdi, B. A. Campbell and J. Ellis, 
Nucl.\ Phys.\ B {\bf 579}, 355 (2000)
[arXiv:hep-ph/0001122].

\bibitem{and}
A. Anisimov and M. Dine, 
Nucl. Phys. B {\bf 619}, 729 (2001)
[arXiv:hep-ph/0008058]. 
A. Anisimov, Phys. Atom. Nucl. {\bf 67}, 640 (2004)
[arXiv:hep-ph/0111233].

\bibitem{kt}
E. W. Kolb and M. S. Turner, {\it The Early Universe},
Addison-Wesley, New York, 1990.

\bibitem{coleman}
S.~R.~Coleman,
Nucl.\ Phys.\ B {\bf 262} (1985) 263
[Erratum-ibid.\ B {\bf 269} (1986) 744].

\bibitem{enqvistetal}
K.~Enqvist and J.~McDonald,
Phys.\ Lett.\ B {\bf 425} (1998) 309
[arXiv:hep-ph/9711514].
K.~Enqvist, A.~Jokinen and J.~McDonald,
Phys.\ Lett.\ B {\bf 483} (2000) 191
[arXiv:hep-ph/0004050].

\bibitem{kusenkoetal}
G.~R.~Dvali, A.~Kusenko and M.~E.~Shaposhnikov,
Phys.\ Lett.\ B {\bf 417} (1998) 99
[arXiv:hep-ph/9707423].
A.~Kusenko and M.~E.~Shaposhnikov,
Phys.\ Lett.\ B {\bf 418} (1998) 46
[arXiv:hep-ph/9709492].

\bibitem{cohenetal}
A.~G.~Cohen, S.~R.~Coleman, H.~Georgi and A.~Manohar,
Nucl.\ Phys.\ B {\bf 272} (1986) 301.

\bibitem{Kasuya}
K.~Enqvist, S.~Kasuya and A.~Mazumdar,
Phys.\ Rev.\ Lett.\  {\bf 89} (2002) 091301
[arXiv:hep-ph/0204270].
K.~Enqvist, S.~Kasuya and A.~Mazumdar,
Phys.\ Rev.\ D {\bf 66} (2002) 043505
[arXiv:hep-ph/0206272].

\end{thebibliography}
\end{document}